\documentclass[12pt]{article}

\usepackage{amsmath, amssymb, amsbsy, amsfonts, latexsym, graphicx}
\usepackage{hyperref}
\usepackage{color, array, subfigure}
\usepackage{cite}

\allowdisplaybreaks

\evensidemargin \oddsidemargin

\begin{document}

%%%%%%%%%%%%%%%%%%%%%%%%%%% TITLE PAGE %%%%%%%%%%%%%%%%%%%%%%%%%%%%%%%
\begin{titlepage}

%-------------------- footnote symbol in title page -----------------
\renewcommand{\thefootnote}{\fnsymbol{footnote}}

%----------------------- preprint number & date ---------------------

%\begin{flushright}
%???-??????
%\end{flushright}

%---------------------------- Title ---------------------------------
\vspace{15mm}
\baselineskip 9mm
\begin{center}
  {\Large \bf 
{{Phase Diagram from Nonlinear Interaction between Superconducting Order and Density:  Toward Data-Based Holographic Superconductor}} 
 }
\end{center}

%--------------------- Authors and Addresses ------------------------
\baselineskip 6mm
\vspace{10mm}
\begin{center}
Sejin Kim$^{a}$, Kyung Kiu Kim$^{b}$,  and Yunseok Seo$^{c}$
 \\[10mm] 
 {\sl College of General Education, Kookmin University, Seoul 02707, Korea}
   \\[2mm]
 {\sl Korea Research Network for Theoretical Physics, Seoul 02707, Korea}
   \\[3mm]
  {\tt  ${}^a$sejin817@kookmin.ac.kr, ${}^b$kimkyungkiu@kookmin.ac.kr, ${}^c$yseo@kookmin.ac.kr}
\end{center}

\thispagestyle{empty}

%-------------------------- abstract --------------------------------
%\vfill
\vspace{1cm}
\begin{center}
{\bf Abstract}
\end{center}
\noindent
We address an inverse problem in modeling holographic superconductors. We focus our research on the critical temperature behavior depicted by experiments. We use a physics-informed neural network method to find a mass function $M(F^2)$, which is necessary to understand phase transition behavior. This mass function describes a nonlinear interaction between superconducting order and charge carrier density. We introduce positional embedding layers to improve the learning process in our algorithm, and the Adam optimization is used to predict the critical temperature data via holographic calculation with appropriate accuracy. Consideration of the positional embedding layers is motivated by the transformer model of natural-language processing in the artificial intelligence (AI) field.  We obtain holographic models that reproduce borderlines of the normal and superconducting phases provided by actual data. Our work is the first holographic attempt to match phase transition data quantitatively obtained from experiments. Also, the present work offers a new methodology for data-based holographic models.
\\ [15mm]

\vspace{5mm}
\end{titlepage}

%%%%%%%%%%%%%%%%%%%%%%%%%%%%%%%%%%
\section{Introduction}
%%%%%%%%%%%%%%%%%%%%%%%%%%%%%%%%%%

Nonlinear electrodynamics (NED) has been actively applied to various fields of physics. One of the famous examples is the Heisenberg-Euler action, which is physically regarded as a one-loop effective action of the quantum electrodynamics \cite{Heisenberg:1936nmg}. Also, the representative work of NED is the Born-Infeld theory \cite{Born:1934gh}, which turns out to be a low-energy effective action of open strings \cite{Fradkin:1985qd, Tseytlin:1985kh} (See, {\it e.g.}, \cite{Gibbons:2001gy} for a review). One advantageous property of the Born-Infeld theory is that this can address the divergence of the electric field self-energy of a point particle. This property motivates a lot of research, such as regular black holes \cite{Ayon-Beato:1998hmi, Breton:2003tk, Dymnikova:2004zc, Balart:2014cga, Kruglov:2015yua, Bronnikov:2000vy} and the early universe or accelerating universe in cosmology, {\it e.g.}, \cite{Garcia-Salcedo:2000ujn, Camara:2004ap, Elizalde:2003ku, Novello:2003kh, Alishahiha:2004eh, Novello:2006ng, Vollick:2008dx, Ahn:2009xd, Kruglov:2015fbl, Kruglov:2016cdm}. Moreover, there are many NED applications in holography. We will come back to such works below.

The AdS/CFT correspondence provided a quantum leap for understanding strongly interacting field theory \cite{Maldacena:1997re, Witten:1998qj, Aharony:1999ti}. See, {\it e.g.}, \cite{Horowitz:2006ct, Polchinski:2010hw, Erdmenger:2018xqz} for reviews. Based on this important discovery, the holographic approach for condensed matter theory (CMT) or AdS/CMT has been developed to reveal secrets of quantum matters\footnote{See, {\it e.g.}, \cite{ Sachdev:2010ch, Hartnoll:2016apf, Baggioli:2021xuv, Zaanen:2021llz} for reviews.}. One of the famous topics is the holographic superconductor \cite{Hartnoll:2008vx}\footnote{See \cite{Hartnoll:2008kx, Hartnoll:2009sz, Herzog:2009xv} for reviews.}, which describes $U(1)$-symmetry breaking with nontrivial superconducting order. This holographic model is extended by a momentum relaxation using an axion field\cite{Andrade:2014xca, Kim:2015dna}. Then, unlike a neutral scalar, the charged scalar was proved essential to carry the genuine superfluid density by a direct conductivity calculation \cite{Kim:2016hzi}.

As mentioned earlier, NEDs are regarded as critical ingredients also in holographic models. Numerous studies have been introduced in the context of extension of the holographic superconductor. See, {\it e.g.}, \cite{Jing:2011vz, Jing:2010zp, Gangopadhyay:2012am, Zhao:2012cn, Liu:2015lit, Lai:2015rva, Sheykhi:2016aoi, Sheykhi:2018mzs, Ghotbabadi:2018ahu, Mohammadi:2019ndh, Lai:2021mxa, Seo:2023kyp}. In these works, one can see that the properties of normal and superconducting phases are affected by the nonlinearity of the Maxwell field. Our previous study showed that nonlinearity is related to a dome-shape superconducting region in the phase diagrams \cite{Seo:2023kyp}. The earlier pioneering studies on the superconducting dome are found in  \cite{Gauntlett:2009dn, Gauntlett:2009bh, Kiritsis:2015hoa, Baggioli:2015dwa, Correa:2019ivh, Cai:2020nyd, Zhao:2023qms}.

Furthermore, the resistivity and the Hall angle display strange behavior in the high-$T_c$ superconductor normal phase. To address this issue through holographic approaches \cite{Kiritsis:2016cpm, Blauvelt:2017koq, Cremonini:2018kla, Bi:2021maw}, NEDs are proposed as a theoretically efficient tool. A physically acceptable interpretation of holographic NED models is that the nonlinearity describes the strong interaction among charge carriers in the many-body picture \cite{Baggioli:2016oju}. This may depict the electron-electron interaction in real materials. Treating such a many-body interaction is another no-easy problem beyond the band-theory framework. This holographic idea was applied to explain Mott insulators in \cite{Baggioli:2016oju}. Therefore, we will take a class of NED interactions as an essential building block that affects the phase transition temperature.

On this ground, one may find a holographic model showing various phases consistent with actual data. As a first step toward a data-based holographic superconductor, we focus on the superconducting dome's boundaries in phase diagrams. Therefore, this paper aims to construct holographic models that are quantitatively consistent with phase diagram data. Our work is the first try to encode experimental data for the correct match through holographic methods.

To help understand the logical steps of the present study, one can compare the holographic method to an actual experiment. Building a gravity model and finding a black hole solution can be linked to preparing material. On the other hand, computing physical quantities via a holographic method corresponds to measuring physical quantities by experiments. Therefore, the procedure of the present work can be compared to manufacturing the material according to measurement data, so our task is a typical inverse problem.

We pick specific phase diagrams of high-$T_c$ superconductor materials as target data. We also consider artificial data to test our methodology before targeting the experimental measurement. In addition, theoretical ingredients are the holographic superconductor model with axion field and nonlinear interaction terms between the Maxwell field and complex scalar. These terms describe a nonlinear interaction between the charge carrier density and the superconducting order in the dual field theory. In more detail, our mission is to find a suitable interaction form that gives the correct boundary of the superconducting phase obtained by experiments.

However, it is a challenging job since the phase boundary is sensitive to the form of the interaction. Hence, in the present study, applying AI technology is desirable to advance our work. See \cite{Hashimoto:2018ftp,  Gan:2017nyt, Park:2023slm, Ahn:2024gjf, Ahn:2024lkh, Cai:2024eqa} as previous studies worth referencing for AI applications in holography. More specifically, our study constructs phase diagrams through a machine-learning application. From a technical point of view, we build a neural network and a loss function to find the complex scalar's effective-mass function $M(F^2)$ consistent with actual data. Thus, the undetermined function $M(F^2)$ can be taken as a neural network to reconstruct a holographic superconductor model. Using this neural network method, we can solve the inverse problem. First, we find models consistent with artificial phase borderlines as a test for our machine-learning code. Then, we obtain models matching with the chosen experimental data.  The main result is shown in Figure \ref{fig_SC2}.

In the holographic model perspective, our investigation is an extension of \cite{Seo:2023bdy, Seo:2023kyp}. In the previous work \cite{Seo:2023kyp}, we gave a particular form of the mass function $M(F^2)$ and see how this NED interaction deforms the superconducting region in the phase diagram. To interpret this extended NED interaction, we take the viewpoint of \cite{Baggioli:2016oju}. Even though the model of \cite{Baggioli:2016oju} has a different normal state from ours, we accept their interpretation. Our study clarifies how the strongly interacting charge carriers affect the shape of the superconducting dome region. Therefore, our holographic result suggests that this interaction plays a vital role in raising the critical temperature. Thus, such an interaction can be a key to understanding the high-$T_c$ superconductor. From the bulk point of view, this is described by a dressed effective mass of the complex scalar by the background electric charges via the nonlinear interaction. We will discuss this issue later.

This project is the first step in finding data-based holographic models for the high-$T_c$ superconductor. Although our achieved result matches the phase boundary between normal and superconducting phases with considerable accuracy, more development is needed for the final goal, {\it i.e.}, the data-based holographic superconductor. In particular, the normal phase should show the strange-metallic characteristic. This required property must be imposed in our next project using a suitable holographic model for strange metals. This modification changes the entire structure of the phase diagram. Nevertheless, our AI algorithm method is robust even for this modified case. Therefore, another significant aspect of the present work is that our methodology is a good starting point for constructing data-based holographic quantum matters.

This paper is organized as follows. We explain our holographic model in section 2. In section 3, we explain the background knowledge about AI techniques. In addition, we describe the neural network to find the form of interaction for a given phase diagram. Section 4 provides our results for artificial data and actual phase diagrams obtained by experiments. In section 5, we conclude our result and explain future directions.

%%%%%%%%%%%%%%%%%%%%%%%%%%%%%%%%%%
\section{Relaxed Holographic Superconductor with Nonlinear Interaction}
%%%%%%%%%%%%%%%%%%%%%%%%%%%%%%%%%%

In this section, we propose our model, which can be an essential element in the neural network. The action is a simple form given by
\begin{equation}
S_{B} = \frac{1}{16\pi G} \int d^4 x\sqrt{-g}\left({\cal L}_{RN}-M(F^2)|\varphi|^2\right),
\end{equation}
where
\begin{equation}
{\cal L}_{RN} = R+ \frac{6}{L^2} - \frac{1}{2}\sum_{i=1}^2\left(\partial\chi^{i}\right)^2 - \frac{1}{4}F^2 -|D\varphi|^2\,.
\end{equation}
Here, the linear axion fields, $\chi^i$, account for momentum relaxation, while the complex scalar field $\varphi$ represents the superconducting order. The Maxwell field strength is given by $F=dA$. In addition, $F^2$ denotes $F_{\mu\nu}F^{\mu\nu}$. $M(F^2)$ is an arbitrary function of $F^2$. Also, we require linear stability of the Maxwell field in this function. The field equations derived from the above action are
\begin{align}
&R_{\mu \nu} = \frac{1}{2}g_{\mu \nu }\left({\cal L}_{RN}-M(F^2)|\varphi|^2\right)+\frac{1}{2}\sum_{i=1}^{2} \partial_\mu \chi^i \partial_\nu \chi^i +\frac{1}{2}D_{\mu}\varphi \left(D_{\nu}\varphi\right)^* \nonumber\\
&\quad \quad ~~ -\frac{1}{2}\left(1+4|\varphi|^2M' \right)F_\mu{}^\rho F_\nu{}_\rho, \nonumber\\
&D^2\varphi = M(F^2) \varphi, ~~\nabla^2 \chi^i  = 0, ~~\nabla_\mu  \left(1+4 |\varphi|^2 M'\right) F^{\mu \nu}  = i g \left(\varphi^* D^\nu \varphi - \varphi D^\nu \varphi^* \right),
\end{align}
where $M'$ denotes the derivative of $M(F^2)$ with respect to $F^2$. The covariant derivative is defined as $D_\mu \varphi = \left(\nabla_\mu - i g A_\mu\right)\varphi$, with $g$ representing the charge of the complex scalar field.

Here, we give a few comments on this model. As the introduction mentioned, our main interest is the boundary between the superconducting and normal phases. We didn't modify the Maxwell kinetic term $-\frac{1}{4}F^2$. A nonlinear kinetic term gives rise to interesting properties, such as strange metal or Mott insulator phase outside the superconducting dome \cite{Kiritsis:2016cpm, Blauvelt:2017koq, Cremonini:2018kla, Bi:2021maw, Baggioli:2016oju}. Although these phases are essential for understanding the high-$T_c$ superconductor, we leave them as ingredients for future studies beyond the present work.  Instead, we focus on the interaction term with superconducting order. This term governs the phase transition temperature and behavior near and below the temperatures. Due to this reason, we adopt a specific form of interaction given by $|\varphi|^2M(F^2)$, which is linear in $|\varphi|^2$. This interaction term does not affect the normal phase, but this model becomes an NED in the broken phase.

Another comment is about the extension of this model. One may consider the higher powers of superconducting order $|\varphi|^2$. Such a consideration changes crucially physics in the deep region of the superconducting dome. Our previous work \cite{Seo:2023kyp}  reported that a strange asymptotic Lifshitz scaling shows up in the deep region of the superconducting dome even by a simple form of $M(F^2)$ up to $(F^2)^2$. Studying the effect from higher powers of $|\varphi|^2$ could shed light on clarifying the quantum critical point principle \cite{Sachdev:2011fcc}. Nevertheless, the present study excludes this extension for simplicity. Another possible consideration is taking into account $\tilde{F}^{\mu\nu}F_{\mu\nu}$ in the model. Such a consideration plays a crucial role in the magnetic response, such as the Hall angle. We ignore this dependence just for simplicity.

We proceed by adopting the following ansatz:
\begin{align}\label{Ansatz00}
ds^2 &= - U(r) e^{2w(r)} dt^2 + \frac{r^2}{L^2} \left(dx^2 + dy^2\right) + \frac{dr^2}{U(r)}\,, \nonumber\\
\chi^i &= \kappa \left(x,y\right), \quad A= A_t(r) dt, \quad \varphi = \phi(r)\,,
\end{align}
where $w(r)$ is chosen to vanishes at the boundary ($r\to\infty$). Taking the gauge $\nabla_\mu A^\mu=0$ with vanishing the radial component $A_r$, the scalar field equation becomes a simple form,
\begin{equation}\label{scalar eq}
\left(\nabla^2- m_{\text{eff}}^2 \right)\varphi = 0\,,
\end{equation}
where the effective mass $m_{\text{eff}}$ is given by
\begin{equation}
m_{\text{eff}}^2 \equiv M(F^2)+ g^2 A_\mu A^\mu =M(F^2)+ g^2 \frac{  e^{-2 w} {A_t}^2  }{ U}\,.
\end{equation}
This quantity approaches constant values in the asymptotic boundary and near horizon. The Breitenlohner-Freedman (BF) bound \cite{Breitenlohner:1982bm,Breitenlohner:1982jf} requires $\left(m^2_{\text{eff}} L^2\right)_{r\to\infty}=m^2 L^2>-9/4$ at the asymptotic boundary. Here, we define the bare mass as $m^2 = M(F^2=0)$ for later convenience.

The ansatz (\ref{Ansatz00}) simplifies the equations of motion as
\begin{align}\label{eom00}
&\partial_r\left(r^2e^w U\phi'\right) = \phi \ r^2 e^w \left(M-g^2\frac{  e^{- 2w}{A_t}^2 }{U}\right)\,, \nonumber\\
&\partial_r \left(r^2 e^{-w} {A_t'} \left(1+4 \phi^2 M'\right)\right)= \frac{2 g^2 r^2 e^{-w} {A_t}  \phi^2}{ U} \,, \nonumber\\
&U'= -\frac{1}{2}r\phi^2\left(M+ g^2 \frac{  e^{-2 w} {A_t}^2  }{ U}+4e^{-2 w}  {A_t'}^2 M'\right)\,\nonumber\\
&\quad~~~~-\frac{1}{4} r e^{-2 w} {A_t'}^2-\frac{\kappa ^2 L^2}{2  r}+\frac{3 r}{L^2}-\frac{U \left(r^2 \phi'^2+2\right)}{2 r}\,, \nonumber\\
&w' = \frac{1}{2} r \left(g^2 \frac{e^{-2 w}{A_t}^2  \phi^2}{U^2}+\phi'^2\right).
\end{align}
The axion equation of motion is automatically satisfied using (\ref{Ansatz00}), so the other equations respect the axion as the relaxation parameter $\kappa$. One can see that this system admits an exact solution known as the Reissner-Nordström (RN)-AdS black brane relaxed by the linear-axion field when $\phi=0$. This solution is dual to the metal (normal) phase, while a hairy solution with nonvanishing $\phi(r)$ describes the superconducting phase.

Now, we demonstrate how to determine the phase boundary of a superconducting dome concerning the critical temperature and charge carrier density. In the gauge/gravity duality, the conserved charge density is obtained from the bulk Maxwell equations. This conserved charge plays a role in the charge carrier density from the holographic DC electric conductivity.  In the high $T_C$ cuprate experiment, the doping parameter is used as a control parameter. The role of doping is to add an electron or hole by replacing an atom in the material. Therefore, changing the doping parameter implies changing charge carrier density, and both are proportional. In this sense, we regard charge carrier density $Q$ as a doping parameter\footnote{It is not a unique way to define doping parameters holographically. In previous work\cite{Kiritsis:2015hoa}, for example, they introduce another $U(1)$ gauge field, which corresponds to the doping parameter. In this case, they can introduce charge carriers and doping parameters individually. In other work\cite{Baggioli:2015dwa}, they defined doping parameters as a ratio of two conserved charges; one is for the charge carriers, and the other is for impurities.
}. 

In the case of the high $T_C$ cuprate, the zero doping region is covered by an anti-ferromagnetic insulating phase. This phase is insulating; hence, the charge carrier density should be zero. This insulating phase still exists at the finite doping region; therefore, the charge carrier density and the doping parameter are not exactly proportional. To match the experimental parameter to the model parameter, we have to take care of all dimensionful parameters and carefully find parameter mapping. 
%This charge density is regarded as the doping parameter in holographic literature
%\footnote{The reason is that doping changes the charge density of a material. This is a conventional identification in most of the holography literature. In this convention, the optimal doping relies on the superconducting dome shape.}. 

In the presence of a black hole horizon and a scalar hair, the temperature and entropy density of the dual system are defined as the following horizon values:
\begin{align}\label{TS}
T_H= \frac{1}{4\pi} U'(r_h) e^{w (r_h)}\,,~~~~s_{\text{th}} =\frac{r_h^2}{4GL^2}\,.
\end{align}
On the other hand, the charge carrier density is defined at the asymptotic boundary. The explicit form is given by
\begin{align}\label{QQ}
Q\equiv\lim_{r\rightarrow \infty}r^2 e^{-w} {A_t'} \left(1+4 \phi^2 M'\right)=\lim_{r\rightarrow \infty}r^2 {A_t'}\,.
\end{align}
We used the asymptotic behavior of the fields in the last equality.

The near-horizon behavior satisfying the horizon regularity can be found as
\begin{align}
&U(r)= 4\pi T_H e^{-w_h}(r-r_h) +\cdots\,,\nonumber\\
&A_t(r) = a_h(r-r_h) +\cdots \, ,\\
&\phi(r) = \phi_h +\frac{ \phi_h M_h}{4\pi T_H e^{-w_h}}(r-r_h)+\cdots\, ,\nonumber\\
&w(r)= w_h +\frac{1}{2}\left(1+\frac{g^2 a_h^2}{M_h^2} \right)\phi'(r_h)^2(r-r_h)+\cdots,
\end{align}
where $w_h=w(r_h)$ and $M_h$ is defined as $M_h \equiv M \left( \left.F^2\right|_{r=r_h}\right)$. $F^2$ is a constant, $-2 a_h^2 e^{-2w_h}$, at the horizon. The parameter set is summarized as follows. $w_h$ is chosen as $w(\infty)=0$, and $T_H$ is determined by the third equation of (\ref{eom00}) in terms of the other parameters. In addition, the horizon location can be taken as $r_h=1$ by scaling the radial coordinate $r$. As a result, only two parameters, $\phi_h$ and $a_h$, remain free.

The RN-AdS black brane with the axion corresponds to the metal(normal) phase, where the $\phi_h$ parameter vanishes. In contrast, the superconducting phase is dual to the hairy black brane with a nonvanishing $\phi_h$. Moreover, we require this scalar hair to be a normalizable mode of the bulk. This situation means the operator dual to the scalar field spontaneously has a nonvanishing expectation value. This normalizability is guaranteed by the asymptotic form of the scalar field as follows:
\begin{align}
    &\phi(r) \approx \frac{J}{r^{\Delta_-}}+\frac{\left<{\cal O}\right>}{r^{\Delta_+}}+\cdots,
    & \Delta_\pm = \frac{3}{2}\pm\sqrt{\frac{9}{4}-m^2 L^2},
    \label{sol_phi}
\end{align}
where $\Delta_{+}$ is the conformal dimension of the dual operator. As is well known, the coefficients $J$ and $\left<\mathcal{O}\right>$ are the dual operator's source and the operator expectation value. The normalizability demands $J=0$ except for some exceptional cases.

Thus, the numerical problem of finding a hairy black hole via spontaneous symmetry breaking is equivalent to finding a solution with $J=0$. As we discussed, our numerical problem depends on the two parameters $\phi_h$ and $a_h$. Furthermore, $J$ relies on the function $M(F^2)$. Therefore, we demand the hairy solutions satisfying the sourceless functional condition, 
\begin{align}
J\left[M(F^2); \phi_h, a_h\right]=0\,.
\end{align}
Near the phase boundary, the linearized equation of (\ref{scalar eq}) for $\varphi$ determines the borderline of the two phases using the above condition. On the other hand, we must be concerned about the back reaction beyond the linearized solution to see the deep inside of the superconducting dome. To check the existence of the back-reacted solutions, we will provide relevant examples of free energy and condensation.

%%%%%%%%%%%%%%%%%%%%%%%%%%%%%%%%%%
\section{Neural Network}
%%%%%%%%%%%%%%%%%%%%%%%%%%%%%%%%%%

%%%%%%%%%%%%%%%%%%%%%%%%%%%%%%%%%%
\subsection{Probe limit near critical temperature}
%%%%%%%%%%%%%%%%%%%%%%%%%%%%%%%%%%

Our previous work \cite{Seo:2023kyp} provides a similar phase diagram of a gravity model with those of the high-$T_c$ superconductors. This work develops the gravity model by introducing an interaction class between the scalar field and higher derivatives of the Maxwell field in the form of $|\varphi|^2M(F^2)$, which can produce various superconducting domes. Building on this, our research aims to solve an inverse problem of finding relevant interaction function $M(F^2)$ for given data of superconductors. This problem is analogous to finding Newton's gravitational potential from the movements of planets.

We adopt the method of Physics-Informed Neural Network (PINN) \cite{Raissi:2017zsi} to solve the above inverse problem. This method enables us to find $M(F^2)$ using neural networks, paying careful attention to adding a positional embedding layer, which would be utilized as a learning device. Positional embedding, a concept used in sequences to sequences and transformer models \cite{Vaswani:2017lxt}, refers to subset data to identify its position and values. This approach adds extra context to machine learning, greatly benefiting the recognition of complex patterns and relationships.

We use boundary information of scalar fields in the black brane geometry, attributed to various $M(F^2)$. The probe approximation to the scalar field near the phase transition is desirable for constructing a phase diagram. We fix the background geometry to be an RN-AdS black brane and turn on the linear fluctuation of the scalar field. This enables us to get the set of parameters where the spontaneous condensation of the scalar field appears. At this step, we apply our machine-learning code to find the mass function consistent with phase diagram data. Also, we check that the resultant mass function reproduces the phase boundaries, and the full backreaction will be considered. In the probe limit, we deal with the charged scalar on the RN-AdS black brane geometry, and the scalar field obeys the following equation:
\begin{align}\label{scalar eq RN}
\partial_r\left(r^2 U\phi'\right) = \phi \ r^2  \left(M(F^2)-g^2\frac{ {A_t}^2 }{U}\right),
\end{align}
where $U$ and $A_t$ are specified as
\begin{align}
U = \left(\frac{r}{r_h}-1\right) \left(\frac{r_h^2 }{L^2}\left(\frac{r}{r_h}+\frac{r_h}{r}+1\right)-\frac{\kappa^2 L^2 r_h}{2 r}-\frac{Q^2}{4 r^2}\right)~,~A_t = Q\left( 1- \frac{r_h}{r}\right)\,.
\end{align}
Then, the scalar equation (\ref{scalar eq RN}) is governed by two parameters, $\kappa$ and $Q$, representing momentum relaxation and charge density in the dual field theory. The square of the field strength $F^2$ becomes $F^2 = -2 Q^2/r^4$ in this probe approximation. The boundary condition of the scalar field is required to obey the given data of critical temperature and charge density. The supposed critical temperature on the phase boundary can be written as
\begin{align}
T_c = \frac{1}{4\pi r_h }\left( \frac{3 r_h^2}{L^2}- \frac{\kappa^2 L^2}{2} - \frac{Q^2}{4 r_h^2}\right).
\label{eq_Tn}
\end{align}
This expression helps pick accessible data points and lets us determine the values of $Q$ and $\kappa$. The gauge coupling $g$ and AdS radius $L$ will be set to 1 in the numerical calculation.

%%%%%%%%%%%%%%%%%%%%%%%%%%%%%%%%%%
\subsection{Physics-informed neural network}
%%%%%%%%%%%%%%%%%%%%%%%%%%%%%%%%%%
\begin{figure}
\centering
\includegraphics[width=0.9\textwidth]{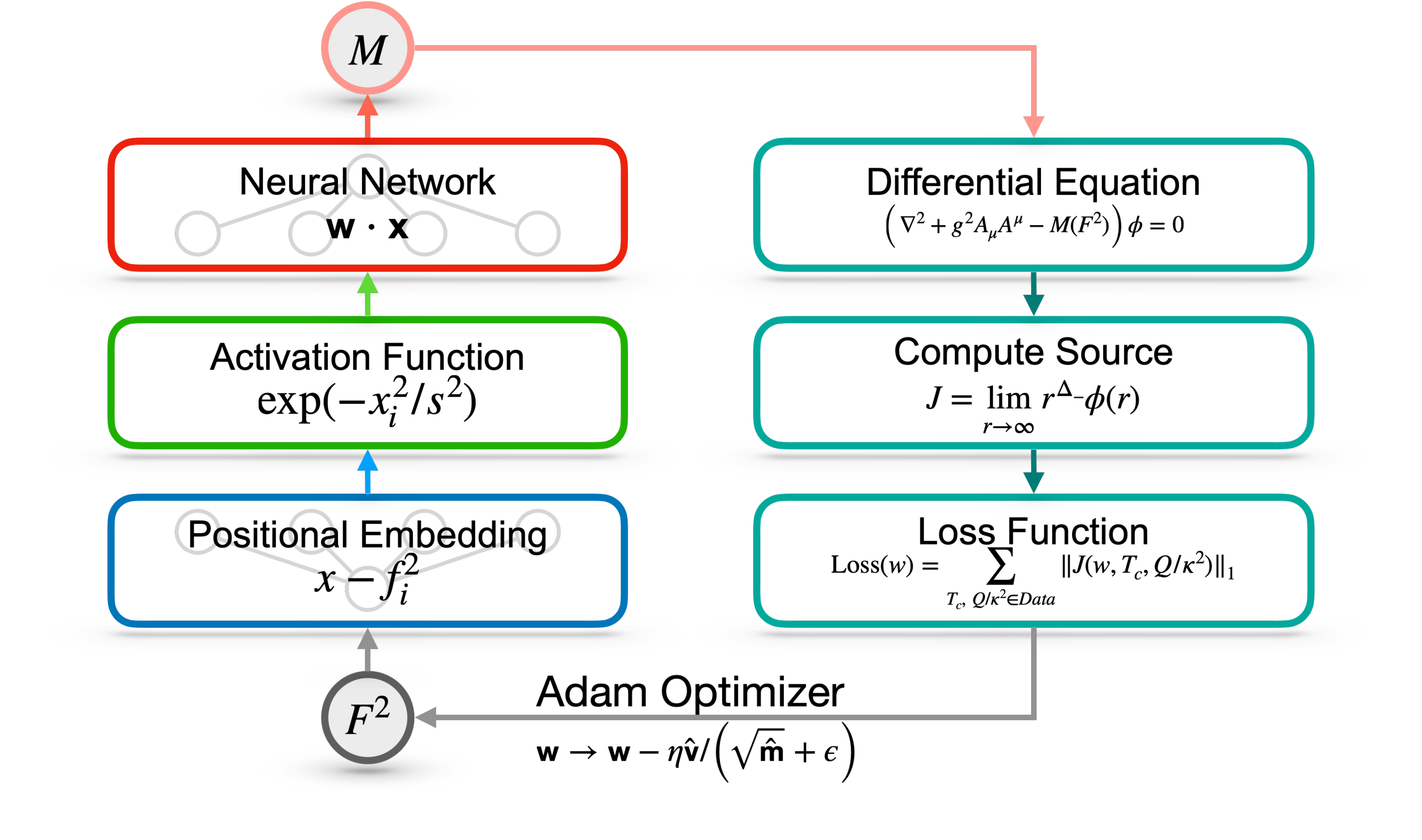}
\caption{This figure illustrates the structured layout of our neural network. The process showcases how the network learns to approximate the mass function 
$M$ from critical temperature data, employing physics-informed strategies for an enhanced understanding of holographic superconductivity.}
\label{NN}
\end{figure}

This section demonstrates how our method uses neural networks to find the inverse problem answer. The most important part of our method is based on the Physics-Informed Neural Networks (PINNs) \cite{Raissi:2017zsi}. PINNs use a neural network as the form of a given solution ansatz, which is relevant for a given physical situation. This means that PINNs exploit the neural network’s capability of approximating complex functions while enforcing the governing physical laws in the training process. The neural network is, therefore, trained not simply on data but to satisfy the differential equations that characterize physical laws, resulting in data-driven and physically meaningful solutions. This method is particularly effective for solving inverse problems of partial differential equations, a common challenge in analyzing physical phenomena. By incorporating a positional embedding layer, our neural network model is uniquely equipped to address the complex inverse problem posed by the holographic approach.

Similarly, the critical idea of Kolmogorov-Arnold Networks (KANs) is to replace the fixed activation functions used in traditional neural networks with learnable activation functions \cite{Liu:2024swq}. Inspired by the Kolmogorov-Arnold representation theorem, KANs feature learnable univariate functions on their edges (weights) rather than fixed activation functions at their nodes,
\begin{align}
    {\rm activation}(t) =  \sum_i c_i B_i(t)
\end{align}
where $B_i(t)$ is B-splines and $c_i$ is trainable. This approach allows each weight to be a parameterized spline function, making the model more flexible and capable of capturing complex patterns with fewer parameters. By combining the strengths of splines and neural networks, KANs achieve higher accuracy and better interpretability than Multi-Layer Perceptrons (MLPs), particularly in small-scale AI and scientific tasks.

Introducing a neural network designed for a specific holographic inverse problem allows for decoding complex equations describing superconductors and other physical entities. The design of this neural network is crucial for taking sufficient critical temperature data and providing a model representing the proper mass function $M$\footnote{We, for simplicity, denote the mass function by $M$, dropping the argument $F^2$ in this subsection.}. 

The neural network has one input layer and one output layer, with a single middle layer having $N$ training variables. At the network's initialization, determining the optimal magnitude of the weights, $w_{i}$, is critical for accurately approximating the mass function. The neural network is formulated as follows:
\begin{align}
M = \text{NN}\left(\sigma (\text{PE}(F^2))\right) = \sum_{i=0}^{N-1} w_i e^{-(F^2- f_i^2)/s^2}\,.
\label{eq_NN}
\end{align}
The right-hand side of the eq.\eqref{eq_NN} is the same as the KAN model, except the basis functions have been replaced from B-splines to Gaussians. The reason we choose Gaussian basis functions is that we do not fix the value of the bare mass $m^2$   to be zero. B-splines have a value only within a given interval, whereas Gaussians do not, allowing the $m^2$ value to be determined naturally. $\sigma$ is a nonlinear activation function, the positional embedding layer PE is a simple operation, and the neural network operation involves matrix multiplication,
\begin{align}
&\text{PE}(x_{i})=x_{i}-f_i^2, \nonumber\\
&\sigma (x)=e^{-x^{2}/s ^{2}}, \nonumber\\
&\text{NN}(x_{i})=\sum _{i} w_{i}x_{i},
\end{align}
where 
\begin{align}
    f_i^2 ={\rm min}(F^2)+i\frac{{\rm max}(F^2)-{\rm min}(F^2)}{N-1}.
\end{align}
Maximum and minimum values of $F^2$ are determined by the critical data of the superconductors and the eq.\eqref{eq_Tn}. In this paper, we take $N=100$ and $s^2=10$.

Using PE is a choice inspired by PE’s application in transformer models. It allows the neural network to understand the several $M$ values in different positions. The ability to integrate positional information is fundamental to our neural network to interpret complex situations and patterns as those obtained from holography data. This is similar to the problem that transformers solve today but with a particular approach to holographic superconductivity data.

With \textbf{NDSolve}, a function in Mathematica that helps solve differential equations, we will solve this equation and integrate the neural network’s weights and the physical parameters into our predictions. To gauge the accuracy of our model, we introduce a loss function,
\begin{align}\label{loss}
{\rm Loss}({w_i}) = \sum_{T_c , \ Q/\kappa^2 \in {\rm Data}} \lVert J\left( w_i, T_c, Q/\kappa^2\right) \rVert_{1}.
\end{align}
Here, by eq.\eqref{sol_phi}, $J$ can be evaluated from the solution of $\phi$ when $w_i$, $T_c$ and $Q/\kappa^2$ are given. To prevent rapid changes in the trainable parameters, we apply the L1-norm, denoted as $\lVert \ldots \rVert_{1}$, also known as the absolute value function. This regularization technique helps maintain stability and ensure smoother convergence during training. The goal is to minimize this loss function, which leads to satisfying the asymptotic AdS boundary condition ($\phi$ to be zero at $r=\infty$), reflecting the disparity between the neural network’s predictions and the expected theoretical results. In this paper, we randomly select 100 sample pairs of the critical temperature and the doping parameter for each training step from the phase diagram data.

Our optimization strategy is the Adam algorithm \cite{Kingma:2014vow}, a powerful method for adjusting the neural network’s weights to minimize the loss function. The Adam optimizer is known for its efficiency in handling large datasets and complex variable relationships. It updates the weights using the following equations,
\begin{align}
&{\textbf v}_{t} = \beta_1 {\textbf v}_{t-1} + (1 - \beta_1) \nabla_{{\textbf w}} {\rm Loss}, \nonumber\\
&{\textbf m}_{t} = \beta_2 {\textbf m}_{t-1} + (1 - \beta_2) \left(\nabla_{{\textbf w}} {\rm Loss}\right)^2, \nonumber\\
&\hat{\textbf v}_{t} = {\textbf v}_{t} / (1 - \beta_1^t), \nonumber\\
&\hat{\textbf m}_{t} = {\textbf m}_{t} / (1 - \beta_2^t), \nonumber\\
&{\textbf w}_{t+1} = {\textbf w}_{t} - \eta\, \hat{{\textbf v}}_{t} / (\sqrt{\hat{\textbf m}_{t}} + \epsilon).
\end{align}
Here, $\textbf w$ represents vector $w_i$, and $t$ signifies the iteration number in the optimization process, serving as a temporal index that tracks the progression through successive steps. ${\textbf v}_t$ and ${\textbf m}_t$ represent the exponentially moving averages of the gradients and the squared gradients, respectively, with ${\textbf v}_t$ focusing on the direction of the steepest descent. At the same time, ${\textbf m}_t$ adapts the learning rate based on the historical gradient magnitude. The decay rates $\beta_1$ and $\beta_2$ are set to 0.99 and 0.999, respectively, with a learning rate $\eta$ of 0.01 and $\epsilon$ at $10^{-8}$ to ensure numerical stability. We summarize the structure of our neural network in Figure \ref{NN}.

%%%%%%%%%%%%%%%%%%%%%%%%%%%%%%%%        
\section{Results}
%%%%%%%%%%%%%%%%%%%%%%%%%%%%%%%%

We designed a neural network as a $\mathbb{R} \rightarrow \mathbb{R}$ function with a single hidden layer containing 100 training variables for the map from the mass function to the location of the phase boundary. In the language of MLPs, the network incorporates a positional embedding layer to handle the data sequencing inspired by transformer models used in natural language processing. Additionally, in the KAN model, a trainable activation function is utilized. The neural network's performance was evaluated based on its ability to approximate $M(F^2)$ from the critical temperature data of superconductors.

The loss function, representing the difference between the neural network's predictions and the target values, was minimized using the Adam optimization algorithm. The final loss value indicated a good fit between the predicted and actual values, demonstrating the effectiveness of the neural network model.

\begin{figure}[ht!]
\centering
\subfigure[]
{\includegraphics[width=0.32\textwidth]{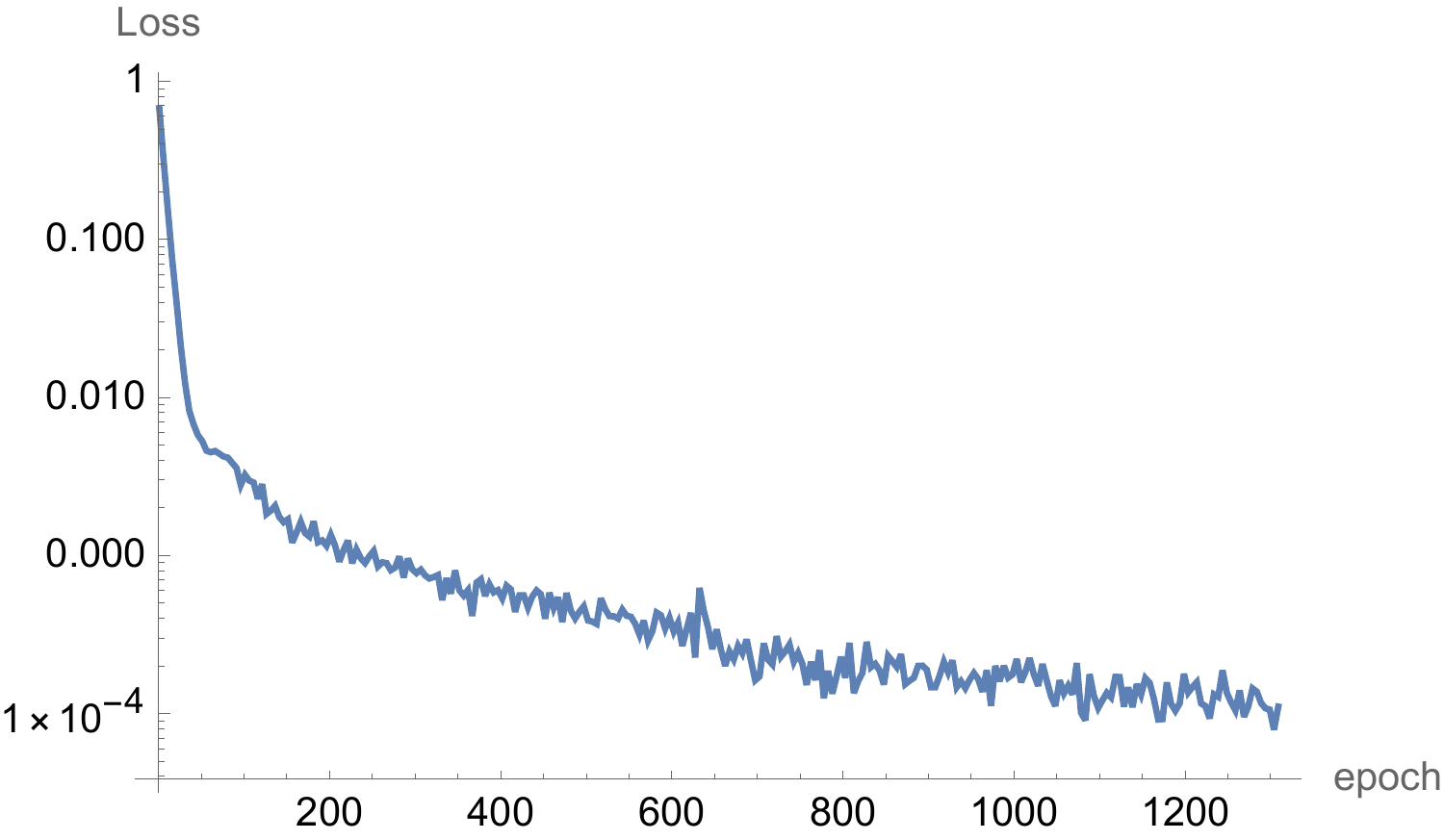}}
\subfigure[]
{\includegraphics[width=0.32\textwidth]{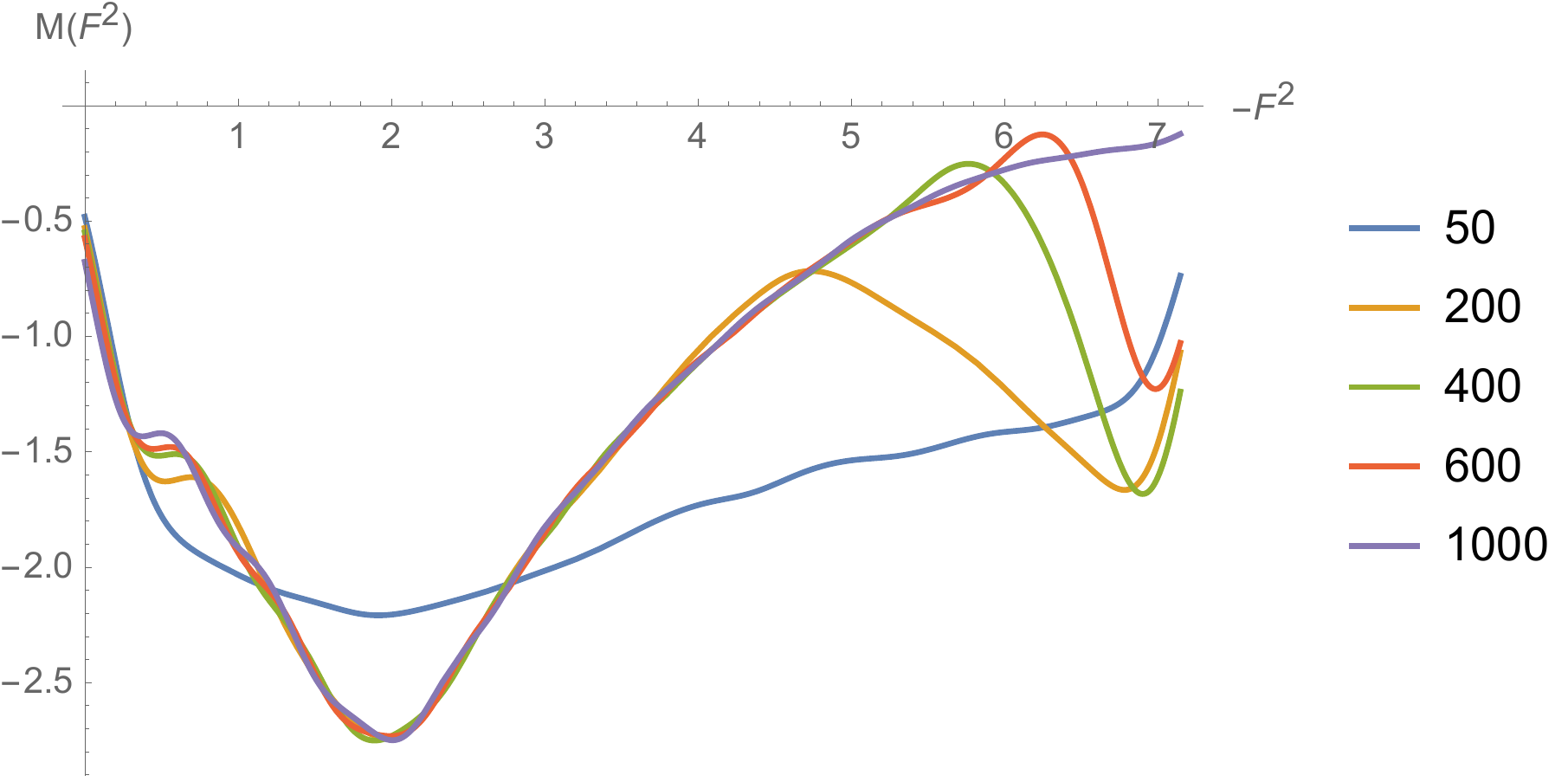}}
\subfigure[]
{\includegraphics[width=0.32\textwidth]{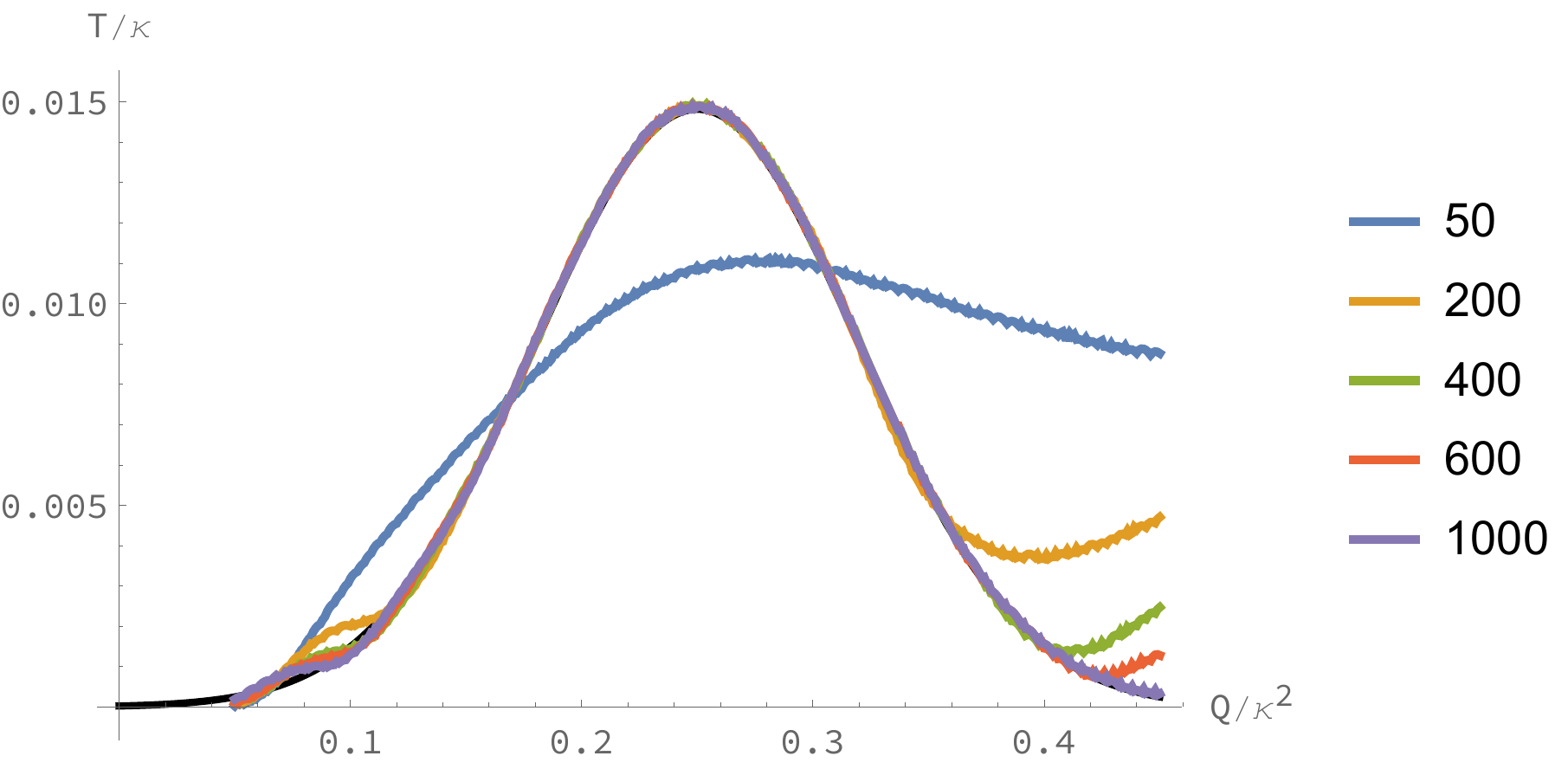}}
\caption{(a) The decrease in the loss function value of our method based on the PINN as training progresses (as epochs increase). (b) Predicting of the mass function $M(F^2)$ as epochs increase as 50, 200, 400, 600, and 1000. (C) The superconducting phase diagram provided by the predicted mass function $M(F^2)$.}
\label{fig_loss}
\end{figure}

The primary goal of our study is to find relevant mass functions $M(F^2)$ for various superconducting phase diagrams. To see efficiency improvement, we visualize the procedures to find a mass function for an artificial Gaussian shape of phase borderline are shown in Figure \ref{fig_loss}. Figure \ref{fig_loss} (a) illustrates the decreasing loss function value of our method as training progresses (as epochs increase). Figure \ref{fig_loss} (b) and (c) show how the mass function $M(F^2)$ is obtained as epochs increase and how the phase diagram is developed to fit the desired shape of the phase diagram finally. As one can see, our neural network models the mass function, successfully generating the target phase borderline.

\begin{figure}[ht!]
\centering
\subfigure[]
{\includegraphics[width=0.35\textwidth]{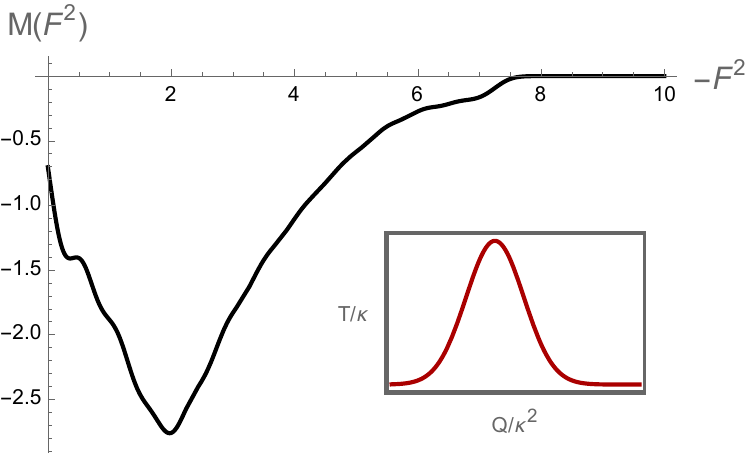}}
\subfigure[]
{\includegraphics[width=0.35\textwidth]{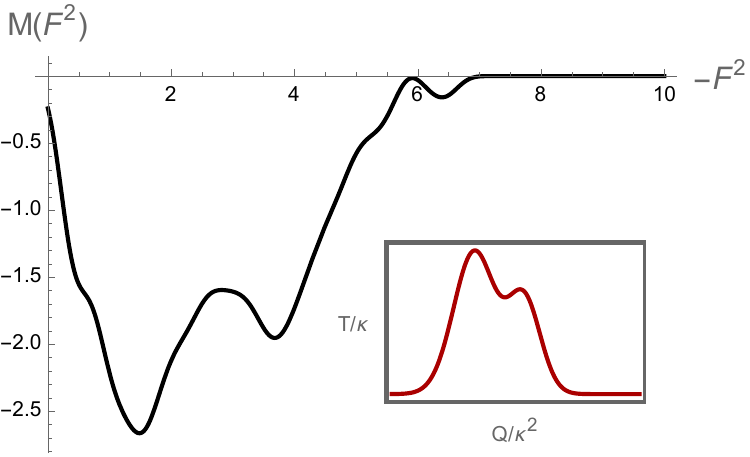}}
\subfigure[]
{\includegraphics[width=0.35\textwidth]{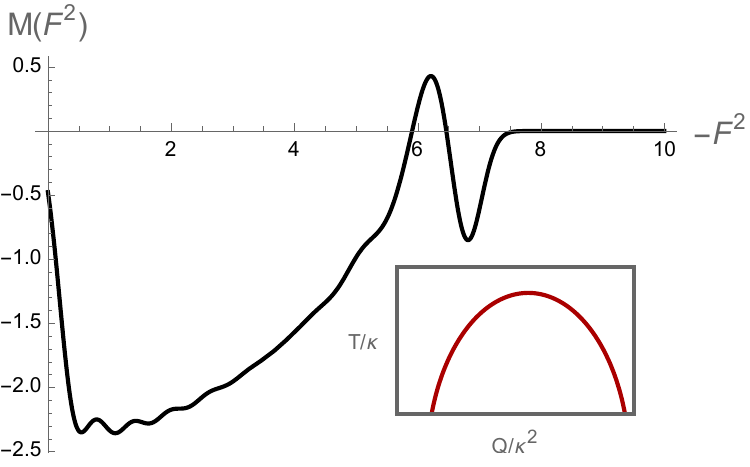}}
\subfigure[]
{\includegraphics[width=0.35\textwidth]{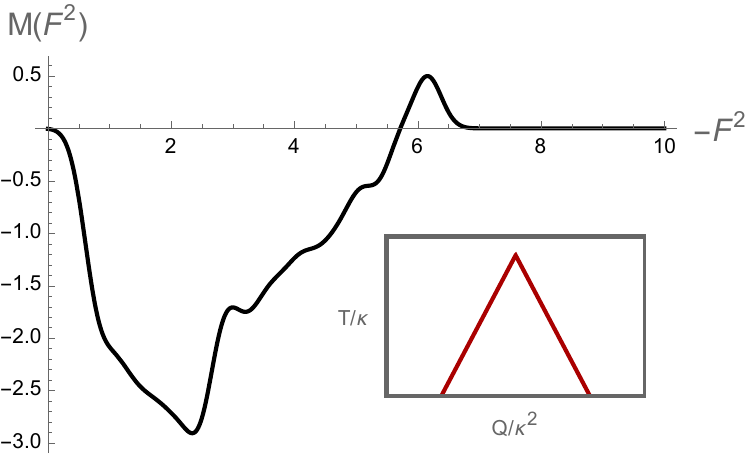}}
\caption{Mass functions for (a) Gaussian Shape, (b) Double Gaussian shape, (c) Semi-circle shape, (d) Triangle shape phase diagram (Inset figure in each figure).}
\label{fig_Mf}
\end{figure}

We apply our PINN code to several artificial shapes of superconducting phase diagrams, such as the Gaussian shape, double Gaussian shape, semi-circle shape, and triangle shape. The mass functions for each phase diagram from the training are shown in Figure \ref{fig_Mf}. As shown in the figure, the mass function $M(F^2)$ can be even positive in some cases, but the condensation of the scalar field can appear. This looks strange from the point of view of BF-bound analysis at the horizon of the extremal black brane. However,  we expect it is due to the non-linear interaction between the scalar field and the $U(1)$ gauge field enough above the zero temperature. There are small wigglings in some mass functions. It comes from the summation of the Gaussian functions with finite width. These wigglings would be flattened if we reduced the width of the Gaussian functions, but we need more training time.

\begin{figure}[ht!]
\centering
\subfigure[]
{\includegraphics[width=0.45\textwidth]{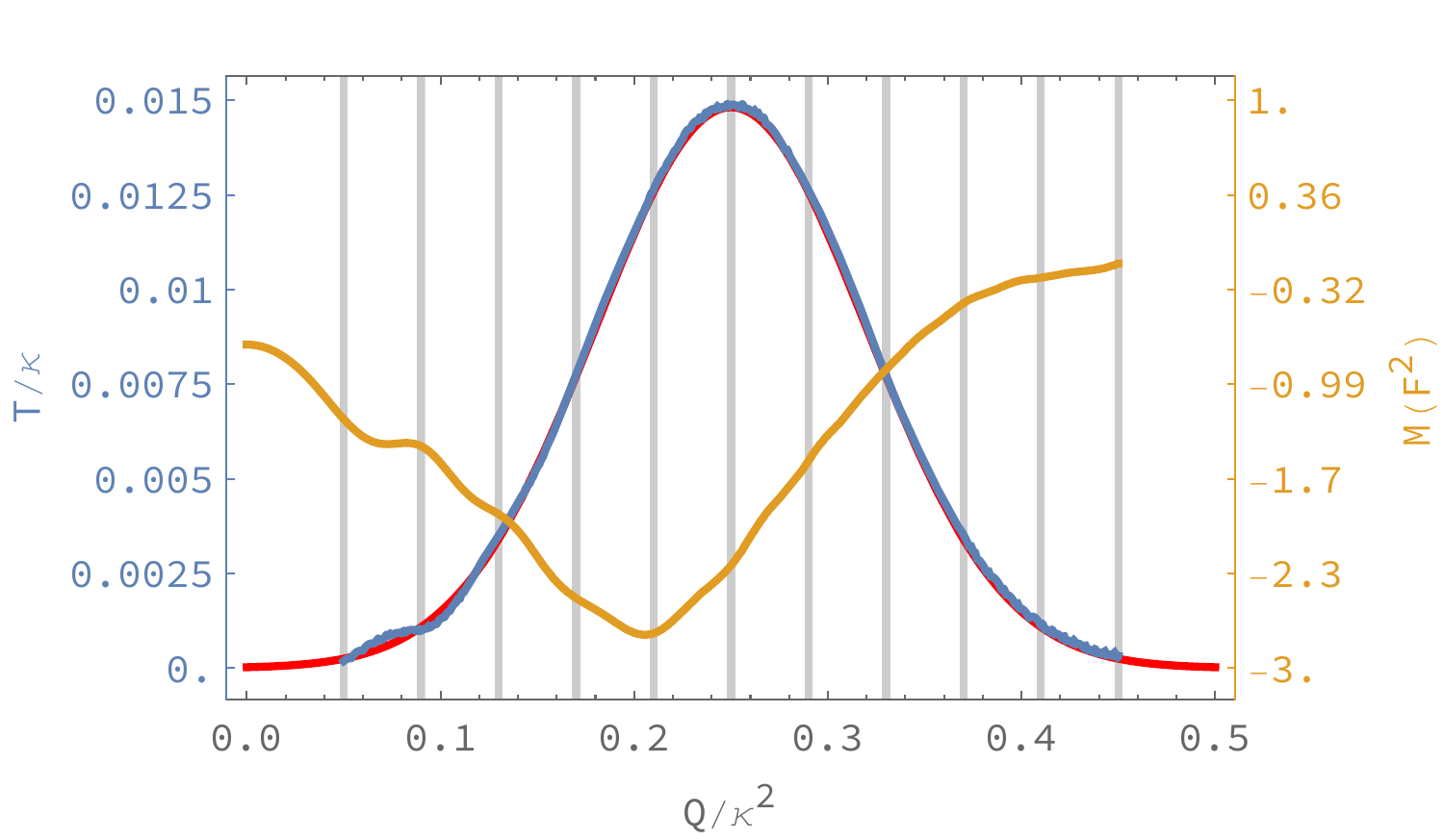}}
\subfigure[]
{\includegraphics[width=0.45\textwidth]{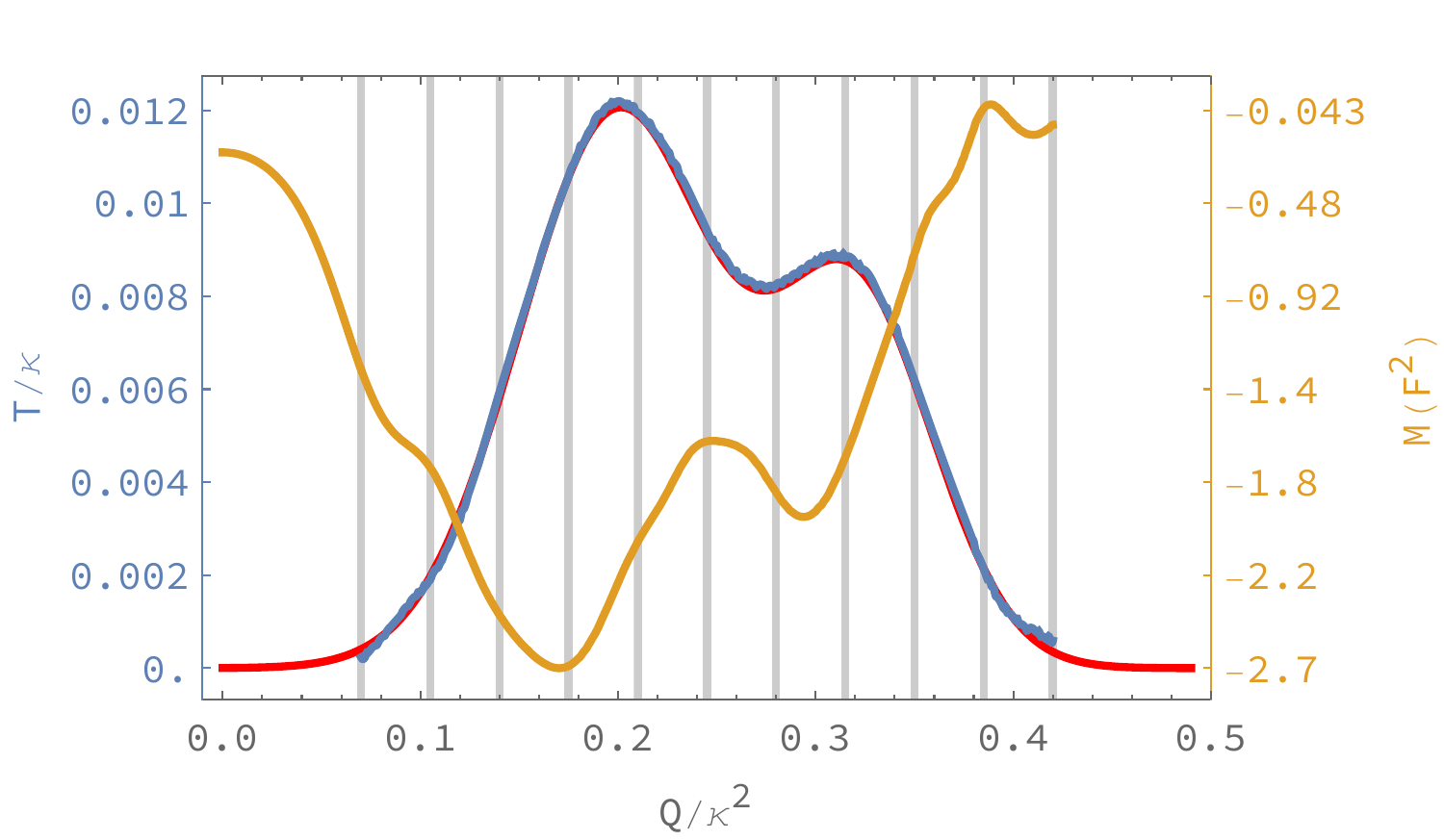}}\\
\subfigure[]
{\includegraphics[width=0.45\textwidth]{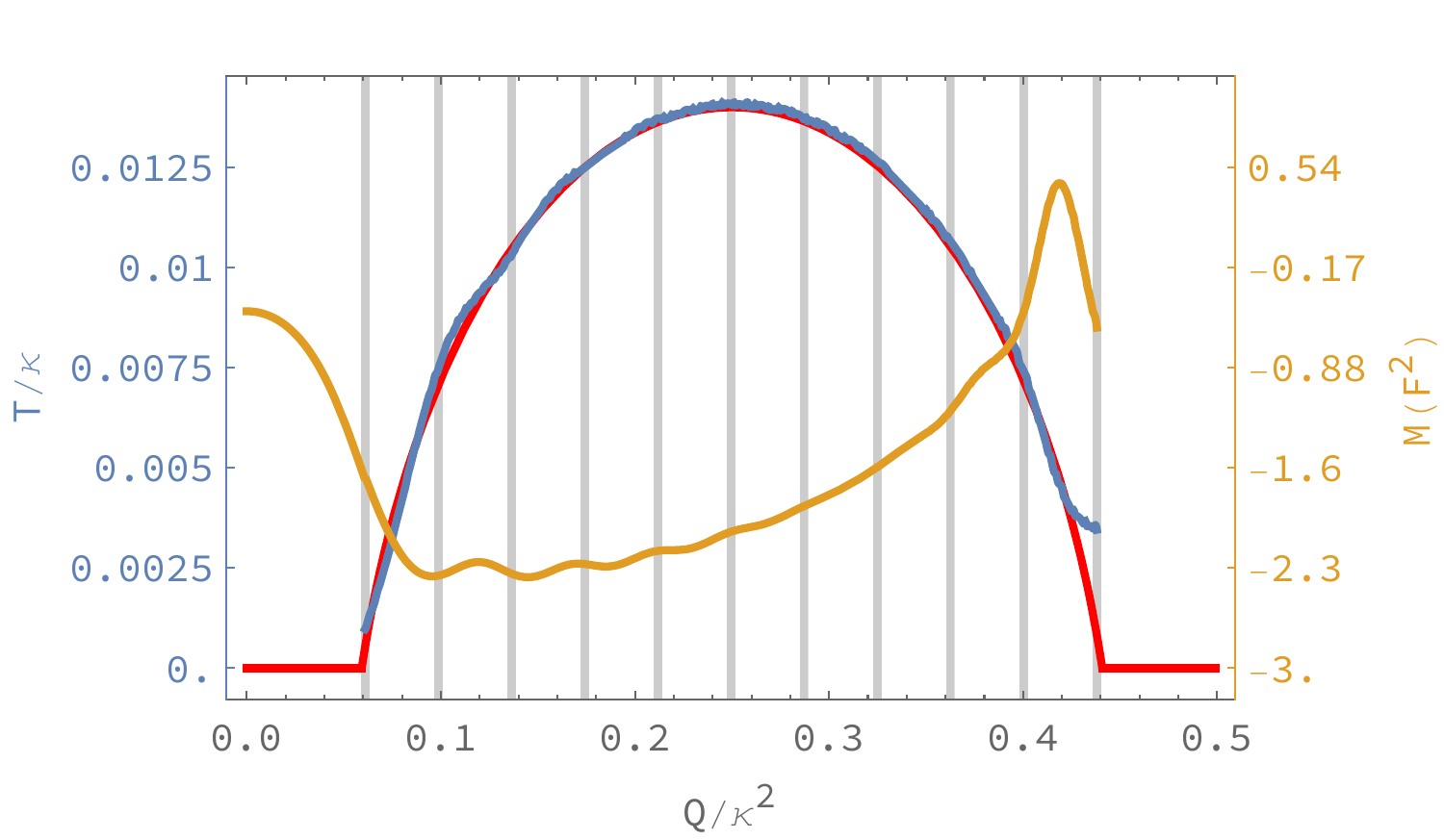}}
\subfigure[]
{\includegraphics[width=0.45\textwidth]{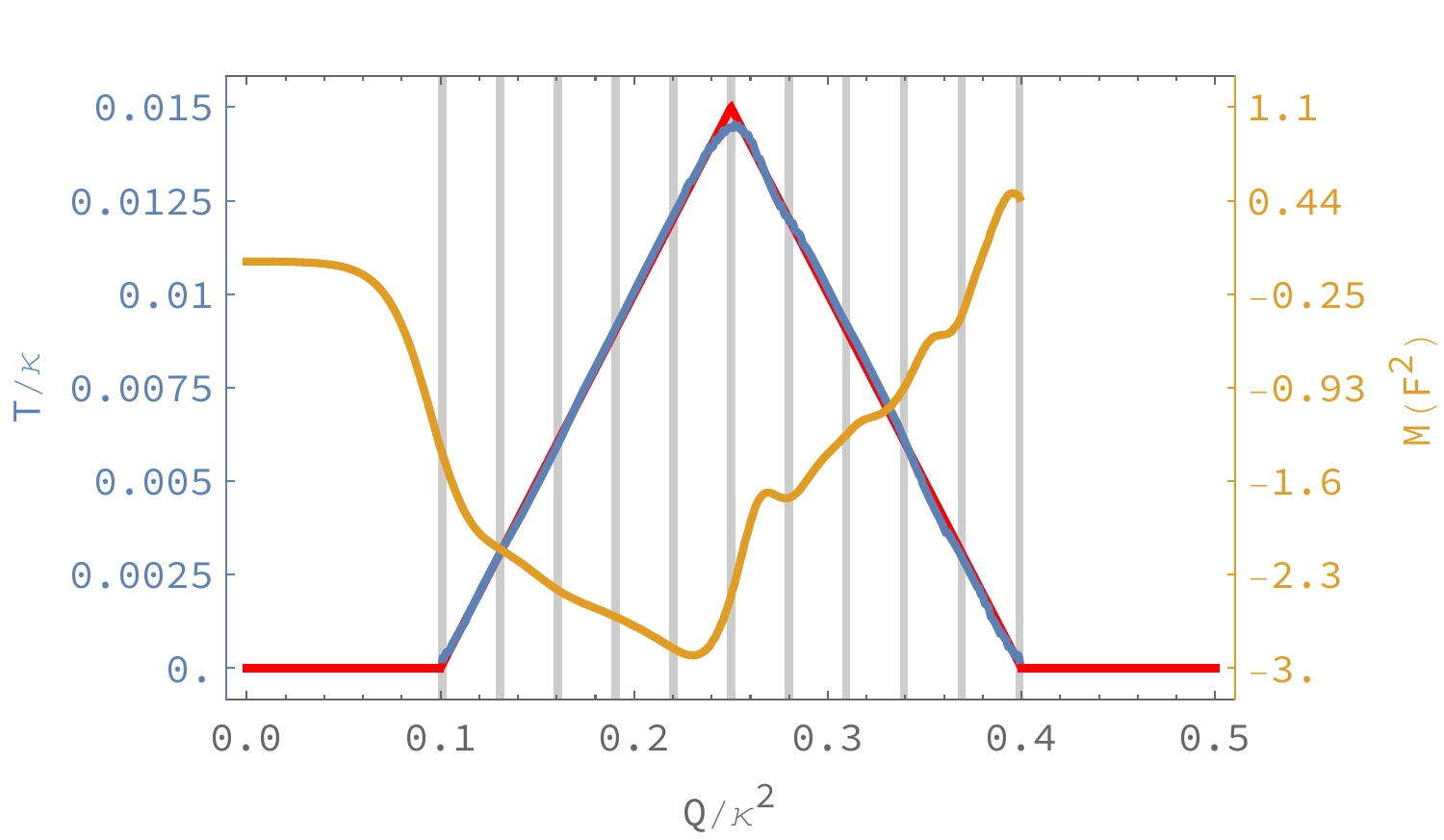}}
\caption{The phase diagram and the mass function  $M(F^2)$ for (a) a Gaussian shape, (b) two overlapping Gaussians, (c) a semicircular shape, and (d) a triangular shape superconductor dome. The achieved loss values (\ref{loss}) are $7.53\times 10^{-5}, 4.81\times 10^{-4}, 2.51\times 10^{-4}$, and $1.07\times 10^{-3}$ for (a), (b), (c), and (d), respectively. The red curves denote the artificial phase borderlines and the blue curves show the borderlines obtained using our PINN method.}
\label{result}
\end{figure}

The mass functions shown in Figure \ref{fig_Mf} are bulk interactions, but the phase diagrams are drawn in the physical parameters of the boundary theory. To see how the mass functions are comparable to the boundary phase diagram, we pick and plot the mass functions $M_h$ at the horizon as a more comparable quantity, given a function of charge density. The summarized results are shown in Figure \ref{result}. The red line in each figure denotes the target (artificial) phase borderlines for the Gaussian, overlapping Gaussian, a semicircle, and the triangular shape superconducting phase diagram. The blue line indicates the phase boundary of the superconducting phase obtained by our neural network method. Two lines match each other well; the orange lines are mass functions at the horizon obtained by our PINN method.

\begin{figure}[ht!]
\centering
\subfigure[]
{\includegraphics[width=0.4\textwidth]{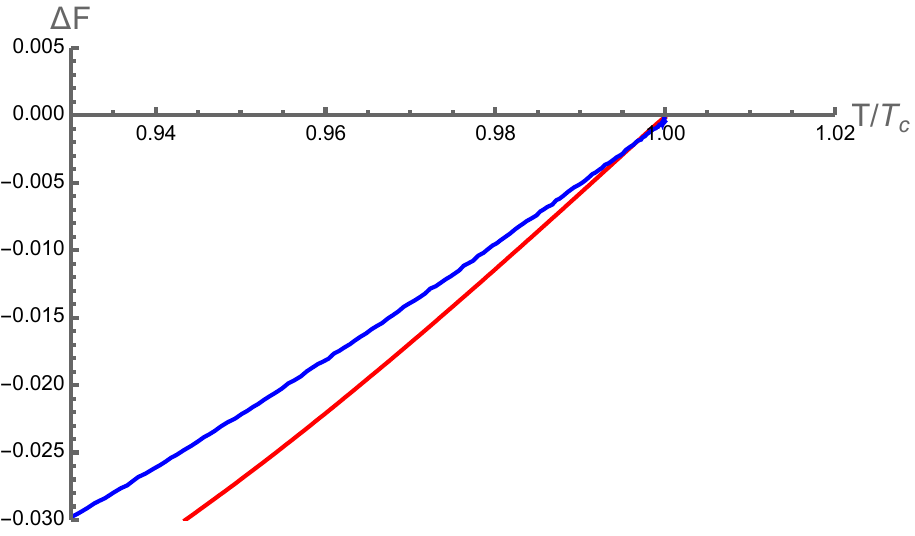}}
\hskip1cm
\subfigure[]
{\includegraphics[width=0.4\textwidth]{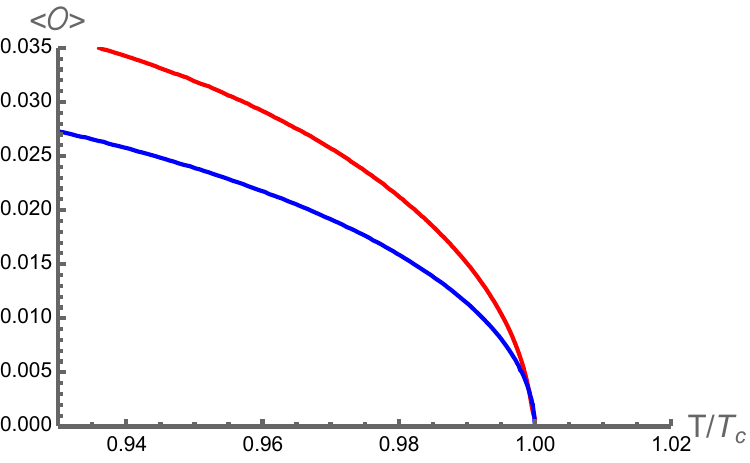}}
\caption{(a) Temperature dependence $\Delta\mathcal{F}$ of the free energy difference for $M(F^2)= -1 +\frac{3}{2}F^2 + \frac{3}{8} (F^2)^2$(red) and $M(F^2) = -4 e^{-(F^2+2)^2/2}$ (blue) to the RN black hole. (b) The condensation curves of the scalar operators for (a).}
\label{fig_Free}
\end{figure}

As a preliminary check of the existence of the hairy solution, including the back reaction, we pick a particular Gaussian shape case and calculate the free energy difference, $\Delta\mathcal{F}=\mathcal{F}_{\text{hairy}}-\mathcal{F}_{RN}$, between its hairy black brane and the RN-AdS black brane. The difference is plotted in Figure \ref{fig_Free} (a). The figure's red line denotes the free energy difference of previous work \cite{Seo:2023kyp}. The blue line shows the present model, where we set $M(F^2) = -4\, e^{-(F^2+2)^2/2}$. In both cases, the free energy of the hairy black brane is always smaller than that of the RN-AdS black brane below the critical temperature; therefore, the hairy solution is energetically preferred at low temperatures. Figure \ref{fig_Free} (b) shows the temperature dependence of the dual operator condensation to the scalar field. The condensation behaves $(T-T_c)^{1/2}$; hence, one can see that the phase transition is a second-order phase transition.

\begin{figure}[ht!]
\centering
\subfigure[]
{\includegraphics[width=0.45\textwidth]{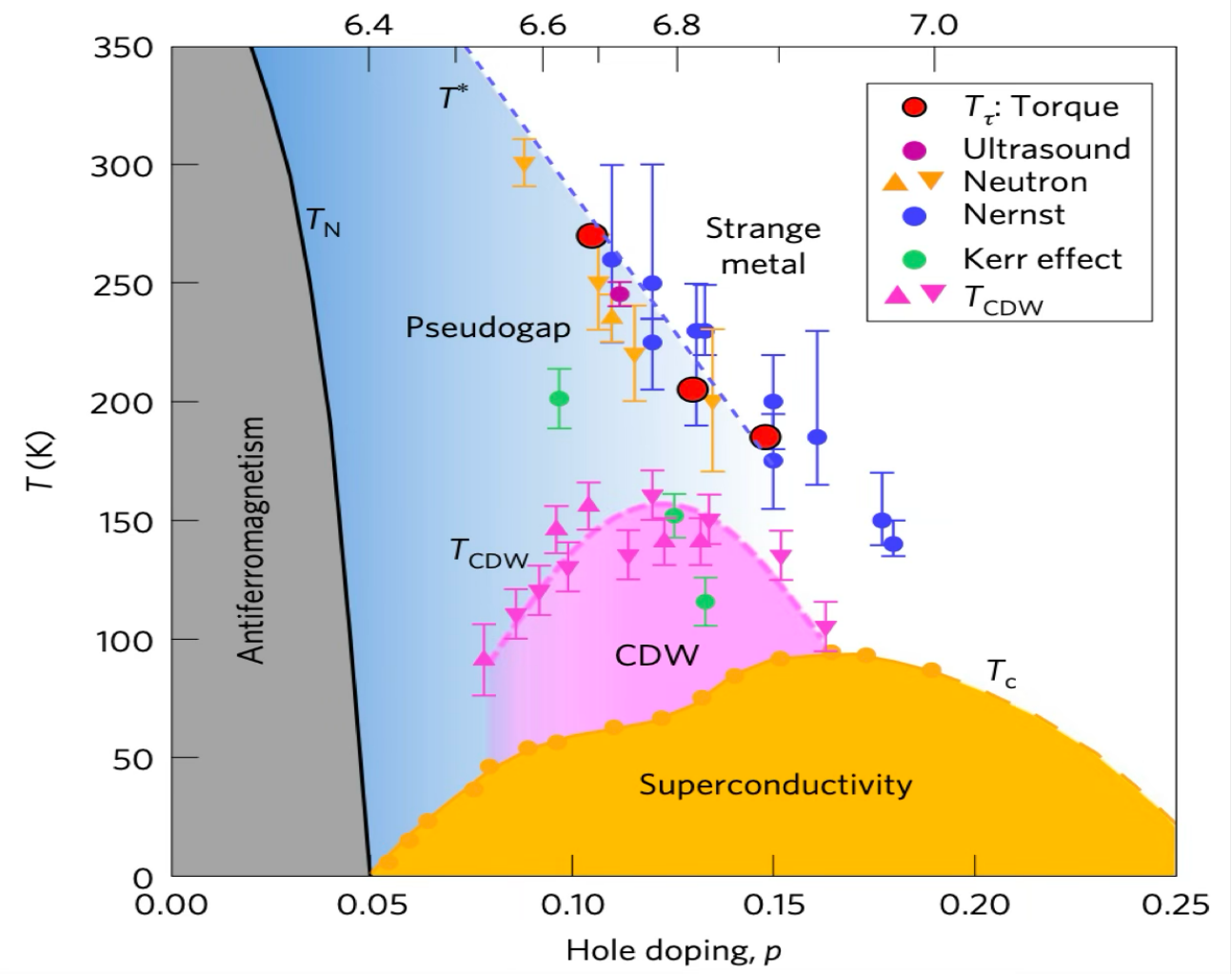}}
\subfigure[]
{\includegraphics[width=0.45\textwidth]{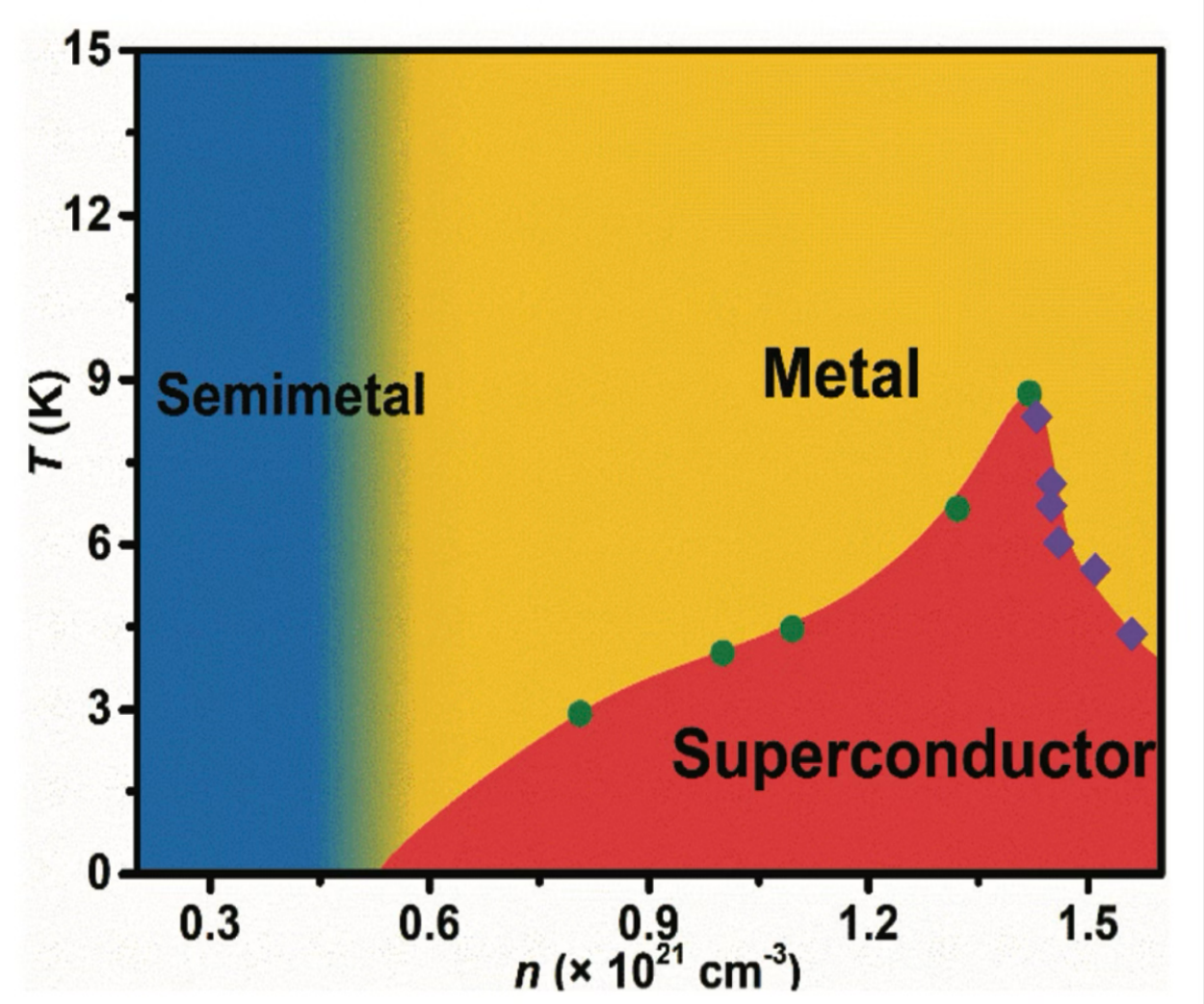}}\\

\caption{The phase diagram of YBa${}_2$Cu${}_3$O${}_\text{y}$ (a) and transition metal dichalcogenies(TMS) (b) superconductor materials. These data were presented in \cite{sato2017, zhao2021p}.}
\label{fig_SC}
\end{figure}

Until now, our PINN method has worked well for finding mass functions for given phase diagrams with various types of shapes. The next task is applying our method to more nontrivial shapes of the superconducting phase diagram or, more explicitly, experimental data of the superconducting materials. We pick up two experimental data sets for the superconducting phase. See Figure \ref{fig_SC}. The figures are superconducting phase diagrams of YBCO (a) and transition metal dichalcogenides (b) superconductors \cite{sato2017, zhao2021p}, which show the non-trivial shapes of the superconducting phase. We apply our method to these cases and find appropriate mass functions.

\begin{figure}[ht!]
\centering
\subfigure[]
{\includegraphics[width=0.45\textwidth]{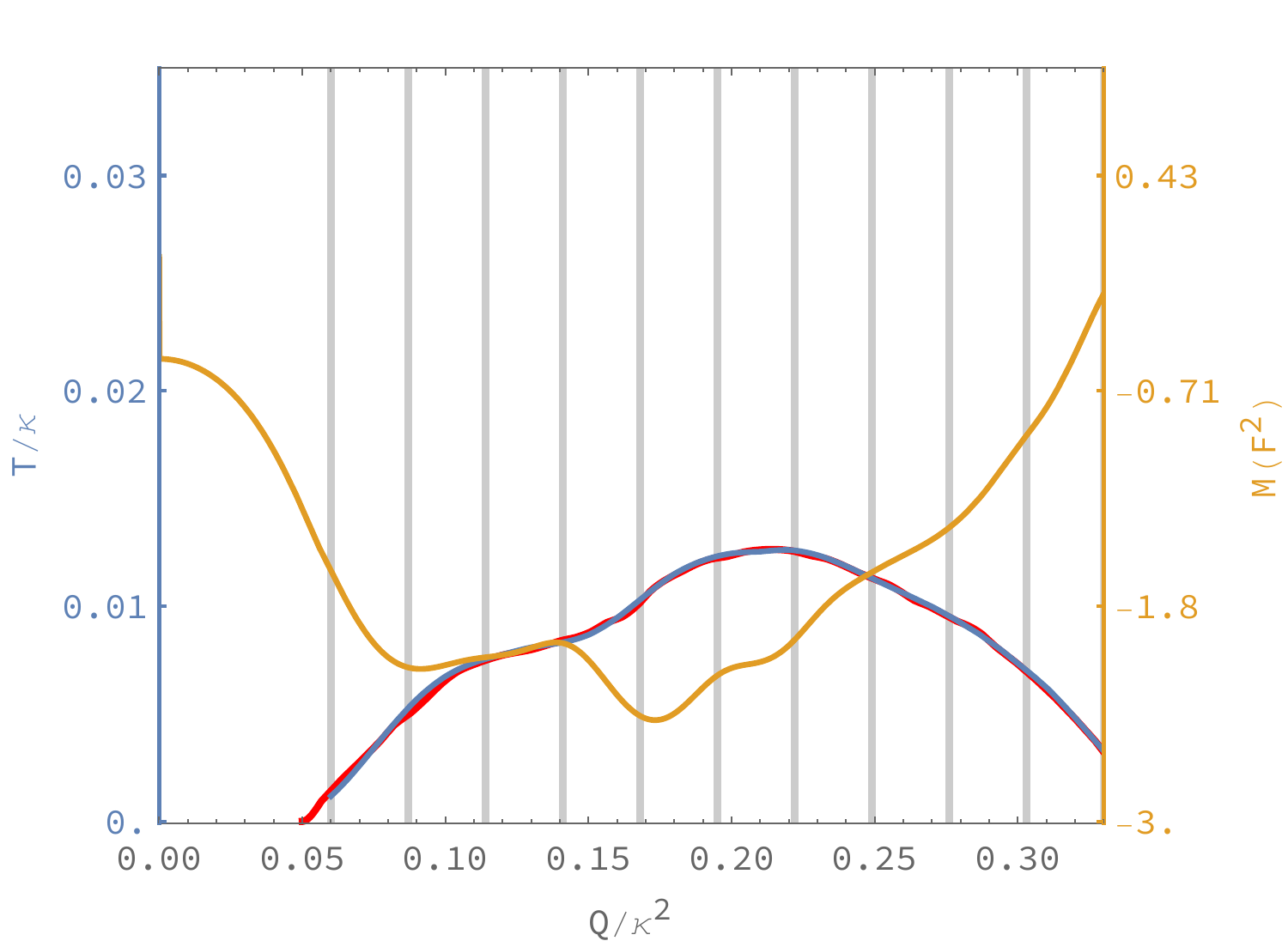}}
\subfigure[]
{\includegraphics[width=0.45\textwidth]{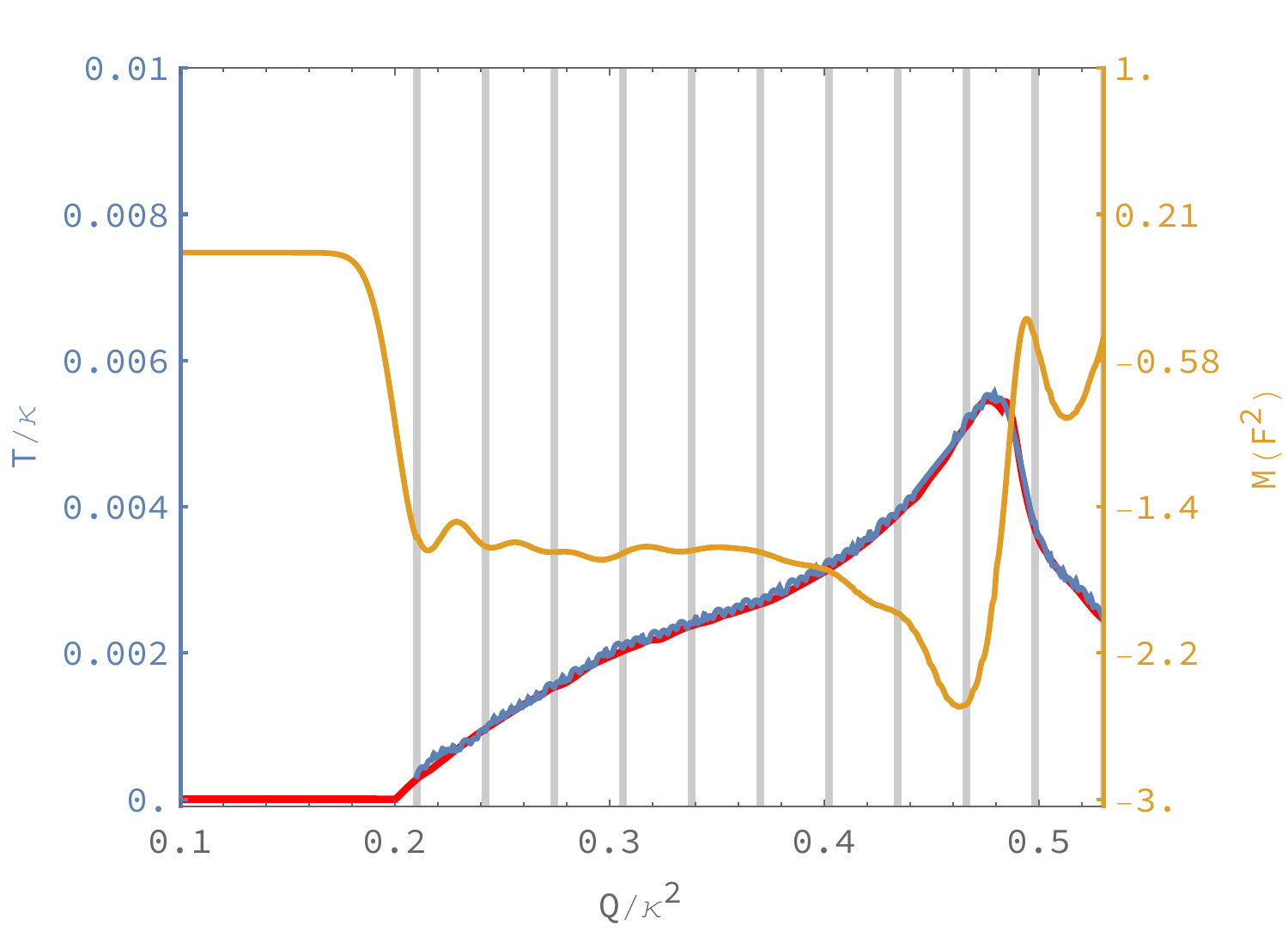}}\\

\caption{Mass function obtained by training(yellow line). The calculated phase diagrams(blue line) match almost the experimental data(red line). The achieved loss values (\ref{loss}) for (a) and (b) are given by $1.64 \times 10^{-4}$  and $8.04\times 10^{-4}$, respectively.}
\label{fig_SC2}
\end{figure}

After the PINN training, we get phase diagrams that match the experimental data. See Figure \ref{fig_SC2}. The red lines of the figures denote the phase borderlines given in Figure \ref{fig_SC}. The yellow lines are mass functions obtained by training, and the blue lines are phase boundaries from these mass functions, which match the experimental data very well. In Figure \ref{fig_SC2}, we use a Gaussian function as an activation function as in (\ref{eq_NN}).

Comparing the actual data (Figure \ref{fig_SC}) and our results (Figure \ref{fig_SC2}), one can notice that the numbers look different. However, since the momentum relaxation parameter $\kappa$ values are different for each material, and the doping parameter and the charge carrier density $Q$ are proportional, but the proportionality constant is also different for each material, the discrepancy in the numbers on the axes can be eliminated by choosing these parameters at any time. We express the numbers on the axes as dimensionless quantities to honestly show this material-dependent ambiguity.

From Figure \ref{result} and \ref{fig_SC2}, one can notice that the mass functions are similar to the phase diagram with a flipped sign. The reasons are as follows: The bare mass of the scalar field is $m^2=M(F^2=0)$ at the boundary, which determines the dimension of the scalar operator. This bare mass does not depend on the charge density. On the other hand, mass functions are highly dependent on the charge density near the horizon. As the effective mass at the horizon becomes more negative, it breaks the BF bound more seriously on the near horizon, and hence, the scalar hair is easily formed. This implies that the transition temperature increases as the horizon effective mass tends to a smaller value toward negative. From the nonlinearity perspective, it is natural to conclude that the negative contribution of nonlinearity to the effective mass increases the critical temperature. Therefore, one may deduce that the electron-electron interaction contributes to the pairing mechanism, giving rise to high critical temperatures in superconducting materials. This reasoning can shed light on understanding the mechanism of high-$T_c$ superconductors.

The effective mass obtained by training fits experimental data very well in Figure \ref{fig_SC2}, but the exact form of the effective mass is very complicated. One can notice that the phase diagram consists of two peaks with different positions and widths. We propose a mass function as a combination of two Gaussian functions as
\begin{align}
 M(F^2) = A_1\, e^{\frac{-(F^2 -B_1)^2}{{\Delta_1}^2}}+A_2\, e^{\frac{-(F^2 -B_2)^2}{{\Delta_2}^2}},
\end{align}
where $A_i$ determines the height of the Gaussian functions. $B_i$ and $\Delta_i$ correspond to the position and width of the Gaussian function, respectively. 
If we choose $A_1=\,-2.45,~A_2=\,0.45,~B_1=\,-1.55,~B_2=-1.25,~\Delta_1=\,2,~\Delta_2=\,1/2\sqrt{2}$, the resulting phase diagram is similar to the phase diagram of the YBCO, Figure \ref{fig_SC} (a) and  $A_1=\,-1.65,~A_2=\,-2.1,~B_1=\,-4.5,~B_2=-6.9,~\Delta_1=\,\sqrt{10},~\Delta_2=\,1/\sqrt{13}$ gives similar phase diagram of the transition metal based superconductor, Figure \ref{fig_SC} (b). From the empirical test of the mass function, the mass function can be determined by the peak feature of the phase diagram.

Also, we checked the free energy difference between the hairy solutions in Figure \ref{fig_SC2} and the RN-AdS black brane. The hairy solutions, including the back reaction, have lower free energies and dominate the superconducting region.

%%%%%%%%%%%%%%%%%%%%%%%%%%%%%%%% 
\section{Conclusion and Future Directions}
%%%%%%%%%%%%%%%%%%%%%%%%%%%%%%%%

We studied a class of gravity models, including an NED between the complex scalar and Maxwell field. This is dual to a superconducting system deformed by nonlinear superconducting order and charge density interaction. In the normal phase, this model class admits the RN-AdS black brane. This is optional but a choice for simplicity. Then, we focus on the phase borderline between the normal and superconducting phases. Our model class has enough degrees of freedom to determine the phase borderlines. Such degrees of freedom are encoded in the scalar's effective mass $M(F^2)$ governing the nonlinear interaction with the background Maxwell field. We used PINNs of machine learning to find relevant mass functions. The pairs of borderlines and mass functions were successfully obtained for various artificial borderlines as a test for our methodology.

Then, we tried to find mass functions for two phase-diagram data obtained from experiments.  Our main result is shown in Figure \ref{fig_SC2}. The location of phase borderlines is affected by the NED of the broken phase. This implies that the charge-carrier self-interaction, comparable to electron-electron interaction in real materials, has a significant role in determining the critical temperature of the dual field theory. Type I superconductors commonly have low critical temperatures, whereas high-$T_c$ superconductors have high critical temperatures. In the view of a holographic method based on our models, it is plausible that superconducting domes accompanied by high critical temperature appear due to the charge-carrier interaction with the superconducting order.

On the other hand, from the gravity model perspective, this phenomenon originates from the background charge. Due to the nonlinear interaction between the complex scalar field and the Maxwell field, the scalar field feels its mass with the dressing effect from the background charge. This effective mass becomes bare mass near the asymptotic boundary. However, the dressing mass effect shows up more and more as the scalar field approaches the horizon. Therefore, the scalar field becomes more unstable near the horizon. This helps the scalar hair form. As a result, the critical temperature becomes higher.

Based on our AI combined result, one can conjecture that high critical temperatures of superconductors may be related to the electron-electron interaction dual to the 
complex scalar dressing mass effect. This cannot be captured in the band theory framework, which relies on the single-particle periodic potential. It would be valuable to study this physics in both theories of holography and condensed matter.

Another notable point of the present work is that we have introduced a new perspective for modeling superconducting domes using a combination of holography and PINNs. Employing positional embedding layers enabled the neural network to learn from data sequences more efficiently, resulting in more coherent and comprehensive insights about how high-$T_c$ superconductors act in various scenarios. We have also shown that such machine learning techniques, particularly neural networks, can be successfully utilized to analyze the inverse problem in theoretical physics. The excellent performance of the Adam optimization algorithm at minimizing the loss function demonstrates that such AI-driven techniques can be used to improve and predict physical models with a high degree of accuracy.

Our study is distinguished by the early introduction of the transformer model in AI technology, motivated by the similarity we noticed between the inverse problem in holography theory and natural language processing tasks. To help understand this similarity, one may compare the translation processing from one sentence to another sentence with our physical problem. These two sentences correspond to the mass-function curve and the phase borderline in our problem. We confirmed that the transformer model can effectively infer the holographic inverse problem. However, due to our hardware limitation, we simplified the model by incorporating only the positional embedding layer of the transformer model. This simplification allows us to achieve the mass function even with the less powerful computer equipment. This simplified tactic aligns with the recently published KAN model \cite{Liu:2024swq}. Moving forward, we will develop our codes to optimize AI models capable of inferring various inverse problems of holography.

%----------------------------

%Too technical description

%Integrating PINNs and positional embedding layers proved to be a powerful tool in modeling complex physical phenomena such as the superconducting dome. The neural network successfully approximated the mass function $M(F^2)$ and provided accurate predictions of critical temperatures. The resulting phase diagrams offered a clear and detailed visualization of the superconducting phases, validating the effectiveness of our approach. Our method demonstrated significant improvements in computational efficiency and accuracy over traditional approaches.

As a comment on the expandability of our work, we found consistent models with the phase borderlines from experiments. Then, we may continue to regions far from the borderlines. One consideration is the normal phase. The high-$T_c$ superconductor has the strange-metallic normal phase, whose resistivity and Hall angle scaling behaviors are $T$ and $T^2$. Several models are proposed in holographic studies as candidates for this scaling property. One famous model is the Gubser-Rocha model \cite{Gubser:2009qt}\footnote{See, {\it i.e.}, \cite{Davison:2013txa, Jeong:2021wiu} for discussions in holographic studies.} for the linear resistivity. In addition, this anomalous scaling behavior has been studied in holography literature. See, {\it e.g.}, \cite{Cremonini:2018kla} as a study with NED. The horizon data encode these response observables to the external fields. See \cite{Kim:2015wba} for an example. Using such a structure, one can construct a new class of models consistent with the strange-metallic property. This model class can also be another candidate for the strange metal. Therefore, our next step toward the data-based holographic superconductor will be substituting the present normal phase of the abovementioned models, showing the anomalous scaling property. Our methodology using the PINNs is robust even in this extension. Once the background geometry is constructed, we follow the same procedure in this paper to get the mass function.

In this work, we focus on the mass function, which gives a similar shape to the phase diagram with experimental data. we set Newton's constant $G$  and the AdS radius $L$ to one. And we also rescale every coordinate to the black hole horizon radius. To match our result to the experimental data quantitatively, we have to consider all dimensionful parameters. This work is a starting point for applying machine learning techniques to the gauge/gravity duality and whether the process is working well.  The background geometry for normal phase does not give a strange metallic behavior. Therefore, we postpone the parameter matching process as a future work.

Another future direction would be to investigate the deep inside of the superconducting dome. A considerable proposal exists to understand the strange metallic behavior via the quantum critical point (QCP) \cite{Sachdev:2009ap, Sachdev:2011fcc}. Our model interaction considers only linear terms in the superconducting order $|\varphi|^2$ described by a magnitude of the complex scalar field. As we discussed earlier, we can introduce the higher powers of the complex scalar field; this may shed light on understanding the deep inside of the superconducting dome, including QCP. Of course, the spin density wave phase must be considered shortly.

%As a final remark, we would like to highlight one implication of our work. Applying our model to the condensed matter physics of other scientific areas shall allow us to understand better the holographic principle and give the potential for their implementation. Thus, AI combined with theoretical physics can improve our knowledge and understanding of perplexing issues.

%%%%%%%%%%%%%%%%%%%%%%%%%%%%%%%% 
\section*{Appendix}
%%%%%%%%%%%%%%%%%%%%%%%%%%%%%%%%

%%%%%%%%%%%%%%%%%%%%%%%%%%%%%%%% 
\section*{A. Direct Search Method}
%%%%%%%%%%%%%%%%%%%%%%%%%%%%%%%%

\begin{figure}[ht!]
\centering
\subfigure[]
{\includegraphics[width=0.45\textwidth]{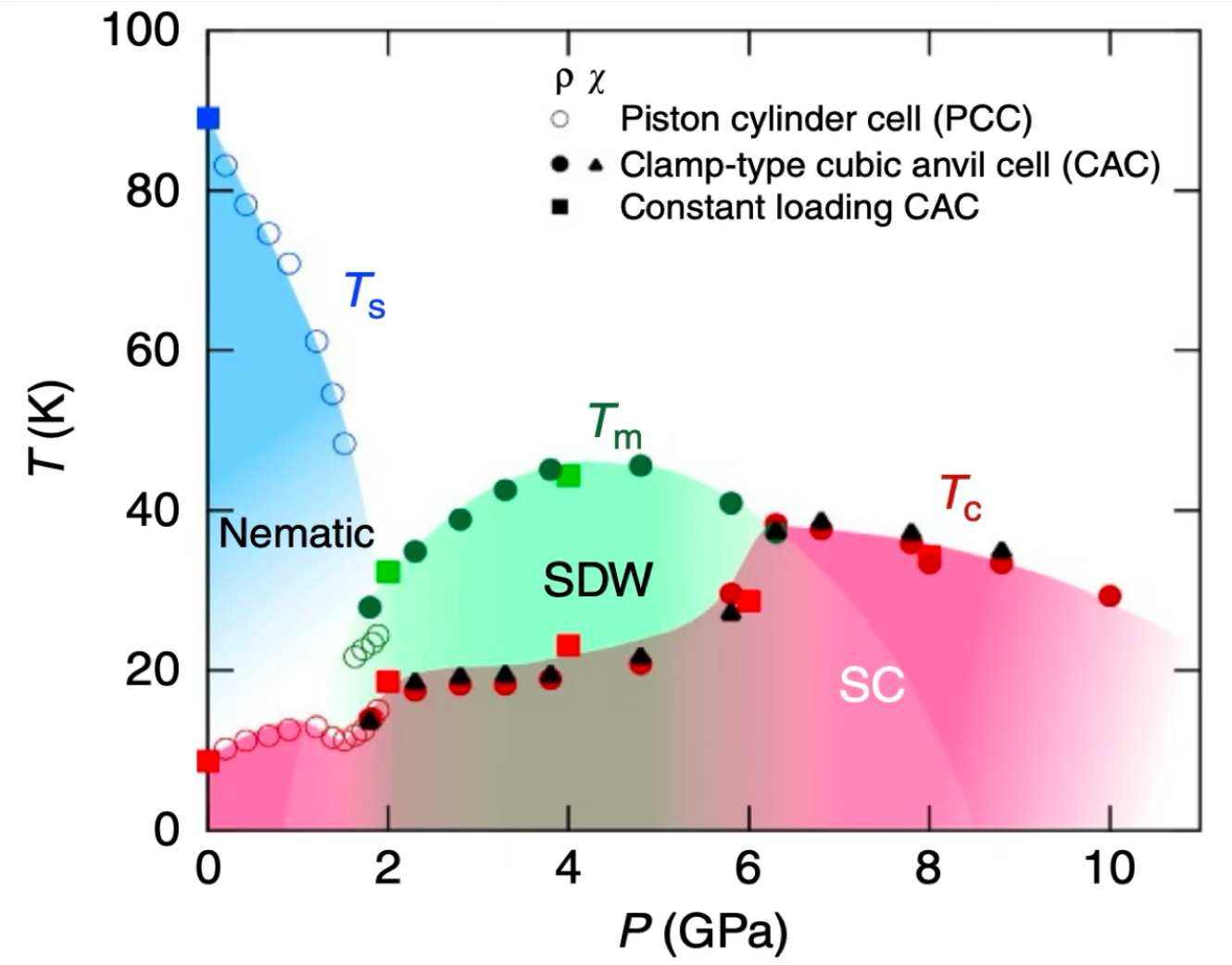}}
\subfigure[]
{\includegraphics[width=0.45\textwidth]{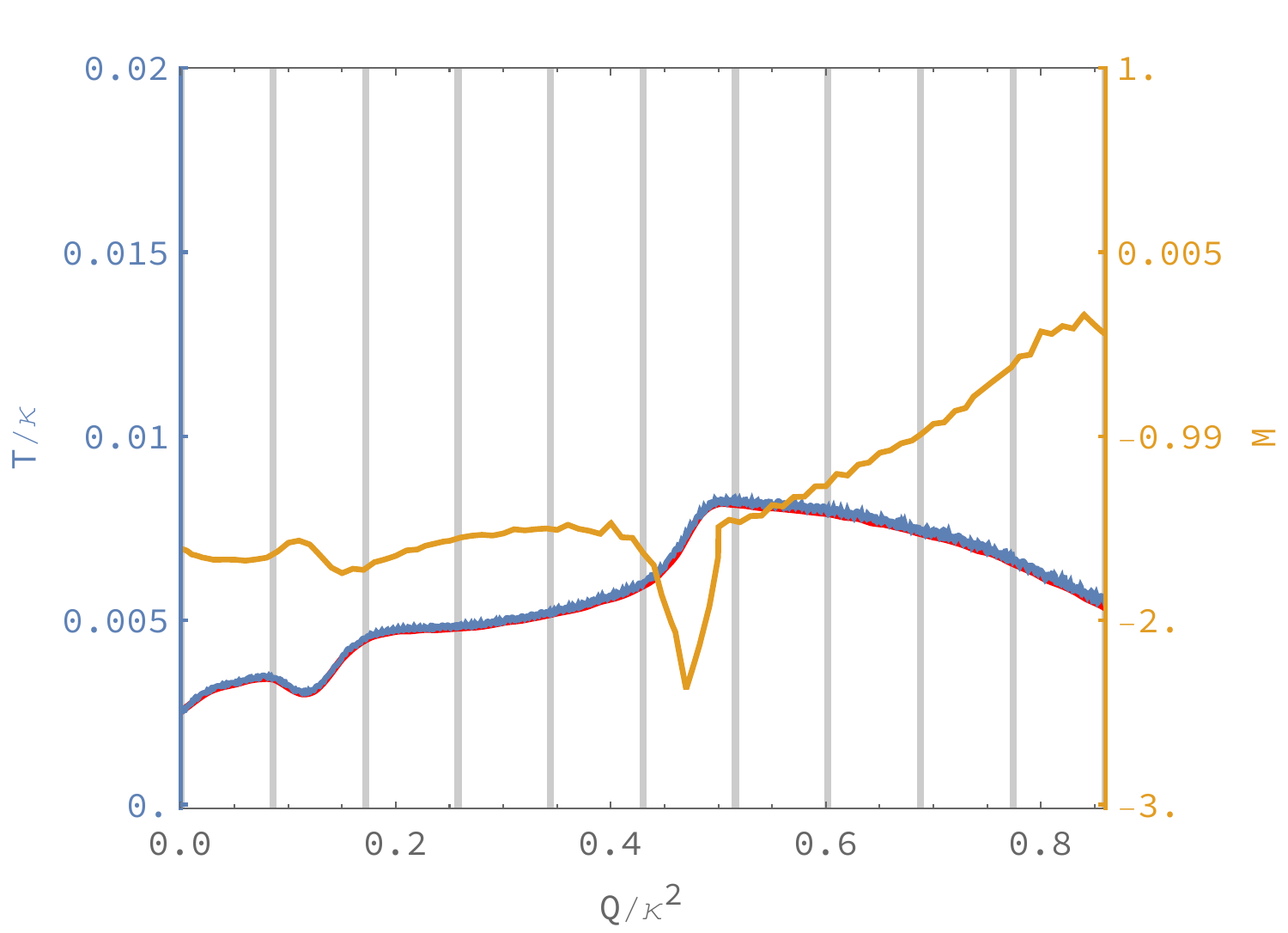}}\\

\caption{(a) The phase diagram of the iron based superconductor materials. These data were presented in \cite{sun2016dome}. (b) shows our result using the direct search method. The loss value of the computation is $1.03\times 10^{-6}$.}
\label{fig_SC-app}
\end{figure}

Here, we want to introduce another effective and relatively simple method of producing a particular shape of the phase diagram. However, this method has one requirement: that the phase borderline should touch the zero-density line. Figure \ref{fig_SC-app} (a) shows an example of this type.

First, we determine the bare mass $m^2=M(F^2=0)$ by solving (\ref{scalar eq RN}) with the vanishing charge density. Then, we extrapolate the function $M(F^2)$ linearly with a small $F^2$. After trying some value $\alpha_1$ of $M(F^2) = m^2 + \alpha_1 F^2$, $\alpha_1$ can be determined by the corresponding phase transition temperature at the small $Q=\delta$. We can continue this extrapolation step by step. In other words, if we accomplished $n$-step with charge density $Q=n \delta$, we can find the $(n+1)$-step coefficient $\alpha_{n+1}$ with $Q=(n+1)\delta$. Using this method, we achieved the result shown in \ref{fig_SC-app} (b). Knowing the starting bare mass is crucial in this method since it determines the operator dimension. Though we need this information, this method is pretty efficient for determining the mass function, and this method can be applied to find other phase borderlines in a whole phase diagram.

%%%%%%%%%%%%%%%%%%%%%%%%%%%%%%%% 
\section*{Acknowledgement}
%%%%%%%%%%%%%%%%%%%%%%%%%%%%%%%%
This work is supported by the Basic Science Research Program through NRF grant
No. NRF- 2022R1A2C1010756(Y. Seo, S. Kim), NRF-2019R1A2C1007396(K. K. Kim).
We acknowledge the hospitality at APCTP, where part of this work was done.

\newpage
%%%%%%%%%%%%%%%%%%%%%%%%%%%%%%%% 
%\section*{Appdendix}
%%%%%%%%%%%%%%%%%%%%%%%%%%%%%%%%

%\bibliographystyle{ieeetr}
%\bibliography{ref}

\end{document}